\documentstyle[12pt]{article}
\newcommand{\sect}[1]{\setcounter{equation}{0}\section{#1}}
\renewcommand{\theequation}{\thesection.\arabic{equation}}
\begin{document}

\begin{flushright}
{\sl 28 October }\\
LPTENS-95/51\\
UT-KOMABA-95/10\\
NSF-ITP-95-151\\
\end{flushright}

\begin{center}
{\large\bf OSCILLATING DENSITY OF STATES
NEAR ZERO ENERGY FOR MATRICES MADE OF BLOCKS WITH
POSSIBLE APPLICATION TO THE RANDOM FLUX PROBLEM}
\end{center}

\begin{center}
{\bf E. Br\'ezin$^{a)}$, S. Hikami$^{b)}$ and A. Zee$^{c)}$}
\end{center}
\vskip 2mm
\begin{center}{$^{a)}$ Laboratoire de Physique
Th\'eorique,{\footnote{ Unit\'e
propre du Centre National de la Recherche Scientifique,
Associ\'ee \`a l'Ecole Normale Sup\'erieure et \`a
l'Universit\'e de
Paris-Sud}} Ecole Normale Sup\'erieure}\\
{24 rue Lhomond, 75231 Paris Cedex 05, France
}\\
{$^{b)}$ Department of Pure and Applied Sciences,
University of
Tokyo}\\
{Meguro-ku, Komaba, Tokyo 153, Japan}\\
{$^{c)}$ Institute for Theoretical Physics}\\
{University of California, Santa Barbara, CA 93106, USA}\\
\end{center}
\vskip 3mm
\begin{abstract}
We consider random hermitian matrices made of complex
blocks. The
symmetries of these matrices force them to have pairs of
opposite real
eigenvalues, so that the average density of eigenvalues
must vanish at the
origin. These densities are studied for finite $N\times N$
matrices in
the Gaussian ensemble. In the large $N$ limit
the density of eigenvalues is given by a semi-circle law.
However, near the
origin there is a region of size $1\over N$ in which this
density rises from zero to the semi-circle, going through an
oscillatory
behavior. This cross-over is calculated explicitly by various
techniques. We
then show to first order in the non-Gaussian character of
the
probability distribution that this oscillatory behavior is
universal, i.e. independent of the probability
distribution. We conjecture that this universality holds to
all orders. We
then extend our consideration to the more complicated
block matrices which arise from lattices of
matrices considered in our previous work. Finally, we study
the case of random real symmetric
matrices made of blocks. By using a remarkable
identity
we are able to determine the oscillatory behavior in this
case
also. The universal oscillations studied here may be
applicable to the problem of a particle propagating on a
lattice with random magnetic flux.
\end{abstract}
\newpage 
\sect{Introduction}

It is well known that the average density of states, for
Gaussian
ensembles
of random $N\times N$ matrices, obeys Wigner's semi-circle
law
when $N$ goes to infinity,
irrespective of the symmetries of the probability measure
\cite{Porter}.
For non-Gaussian measures the average density depends
 sensitively upon the distribution \cite{BIPZ}. However, next
to the edge
of the support of
the eigenvalue
distribution, there is a region of size $N^{-2/3}$, in which
the
average density
crosses over from nonzero to zero, with a universal
cross-over
function (i.e. independent of the probability distribution)
\cite{BK}.
In this work we consider instead an ensemble of random
hermitian
matrices made of complex blocks.  These matrices have been
discussed recently for its application
to impurity scattering in the presence of a magnetic field
\cite{HZ,HSW}
 and to a study of the
zero modes of a Dirac operator\cite{Verbaarschot}.
In the large $N$
limit the  average density of eigenvalues
is again  a semi-circle for Gaussian ensembles. However, by
construction
these matrices have pairs
of opposite real eigenvalues. Thus, as an eigenvalue
approaches zero, the
mid-point of the spectrum, it is repelled by its mirror image.
Consequently, the density of eigenvalues is constrained to
vanish at the origin. Away from the origin, the density
must rise rapidly, over a region
of size
$1\over{N}$, towards the Wigner semi-circle.

In a recent work \cite{HZ} we showed, by explicit
computation in the
Gaussian case, that the rise ``overshoots" the Wigner
semi-circle and thus
has to come back down, whereupon it overshoots again.
Thus, the density of
eigenvalues oscillates over
a region of size $1\over N$. This cross-over at the center of
the spectrum is however not of the same nature as the
cross-over at the end of the spectrum.

 For the simplest case of one random
matrix, we can calculate the cross-over by
three different methods: i) the orthogonal polynomial
approach \cite{Mehta},
 ii) a
method inspired by Kazakov's approach
 to the usual hermitian Gaussian problem \cite{Kazakov},
iii) a supersymmetric method based on Grassmannian
variables \cite{Brgrass}.
The first method is in fact quite cumbersome and hard to
generalize to more
difficult problems. The second one provides an elegant
integral
representation of the correlation functions for finite $N$.
However it is
only through the
Grassmannian approach that we could tackle
the  more complicated problems of a whole lattice of
matrices discussed in our
previous work \cite{BZlattice}.

For the most part the calculations in this paper are done
with the Gaussian distribution. An interesting question is
whether the cross-over behavior at the center of the
spectrum is universal, that is, independent of the details of
the probability distribution. A simple scaling argument
suggests universality. As usual in random
matrix theory \cite{Porter}, the eigenvalues can be thought
of as
representing gas particles in a one-dimensional space. The
probability distribution of the random matrices determine
the potential confining the gas. This confining potential
controls the width of the eigenvalue spectrum, which is of
order $N^0$ in our convention. Crudely speaking, the
oscillatory
behavior we are discussing here depends on the repulsion
between an eigenvalue and its mirror image, and on the
repulsion between the eigenvalues. It
should be possible to regard the confining potential as
essentially constant over the region of size
$1\over N$ we are concerned with and hence irrelevant.
However, further
thought reveals that this argument is insufficient, since the
confining potential, by changing the width of the spectrum,
effectively also changes the average value of the density of
eigenvalues near the center of the spectrum. We also expect
that as the width of the entire spectrum changes the period
of the oscillations near the center of the spectrum would
also
change accordingly. We are able to
show, to first order in the non-Gaussian character of the
distribution, that these effects cancel out, and that the
cross-over behavior is universal. We conjecture that this
cross-over behavior is indeed universal to all orders.

We then show that we can extend our considerations to
study random real
symmetric matrices made of real (but not symmetric)
blocks. By using a
remarkable identity, the universal oscillations near the
center of the
spectrum can again be calculated explicitly.

Finally, we discuss a possible realization of these universal
oscillations
in the problem of a single particle propagating on a square
lattice
penetrated by a random magnetic flux.

\sect{Orthogonal polynomials approach}

In the simple one-matrix problem we consider $2N\times
2N$ block matrices $M$
of the form
\begin{equation}\label{C1}
    M  =\left(\matrix{0& C^{\dag}\cr
                 C& 0 \cr}\right),
\end{equation}
in which $C$ is an  $N\times N$ complex random matrix, with
probability
distribution
\begin{equation}\label{Gaussian}
  P(C) = {1\over{Z}} {\exp} (-N {\rm Tr} C^{\dag}C).
\end{equation}
It is easy to show that $M$ has pairs of opposite real
eigenvalues; indeed if
 $\left(\matrix{x\cr y \cr}\right)$ is an eigenvector for the
eigenvalue
$\lambda $,  then  $\left(\matrix{x\cr -y \cr}\right)$ is an
eigenvector
for  $-\lambda $. In other words the matrix $M$ anti-
commutes
with the $``\gamma_5"$ matrix $\left(\matrix{1& 0\cr
0& -1 \cr}\right)$.
Therefore one can express the average resolvent of the
matrix $M$ in terms of
that of $C^{\dag}C$, a hermitian matrix with positive
eigenvalues:
\begin{eqnarray}\label{C2}
   G(z) &=& <{1\over{2N}} {\rm Tr} {1\over{z -
M}}>\nonumber\\
       &=& <{1\over N}  {\rm Tr}
{z\over{z^2 - C^{\dag}C}}>.
\end{eqnarray}
Taking the imaginary part of (\ref{C2}) we relate the
density of
eigenvalues of $M$
\begin{equation}\label{D1}
  \rho(\lambda) = <{1\over{2N}} {\rm Tr}  \delta(\lambda -M
)>
\end{equation}
to that of $C^{\dag}C$
\begin{equation}\label{D2}
   \tilde \rho(r) = <{1\over{N}} {\rm Tr}  \delta(r- C^{\dag}C)>
\end{equation}
as
\begin{equation}\label{D3}
  \rho(\lambda) = |\lambda |\tilde\rho(\lambda^2).
\end{equation}

If we integrate out as usual over the unitary group, we are
left with
integrals over the $N$ eigenvalues $r_i$ of $C^{\dag}C$,
which run from
zero to infinity. The Jacobian of this change of variables is
simply the
square of the Van der Monde determinant of the  $r_i$'s.
The probability
measure is then
\begin{equation}\label{C3}
  P(r_1,\cdots,r_N) = {1\over{Z}} {\exp} (-N {\sum_1^N} r_i)
\Delta^2(r_1,\cdots,r_N).
\end{equation}
All the correlation functions are known to be expressible in
terms of the kernel
\begin{equation}\label{C4}
  K(r,s) =   {\exp} (-{N(r+s)\over{2}}) \sum_0^{N-1}
L_n(Nr)L_n(Ns).
\end{equation}
in which the polynomials $L_n(x)$ are the Laguerre
polynomials, orthogonal
on the half-line with the measure $\exp(-x)$.  The density of
eigenvalues
of $C^{\dag}C$ is then given by
\begin{equation}\label{C5}
   <{1\over{N}} {\rm Tr} \delta(r - C^{\dag}C)> = K(r,r).
\end{equation}
Using the Christoffel-Darboux identity, and the asymptotic
behavior of
\\$L_n(Nr)$ for $n$ of order $N$,
 we could obtain the desired density of state from this
method. But not only is this method more cumbersome for
this one-matrix
case, also it cannot be generalized to the lattice of matrices
that
we discuss
below. We shall return to it later after we considered the
cross-over near
the origin.
\sect{Kazakov's method extended to complex matrices}
\vskip 5mm
{\bf 3.1 Contour integral}
\vskip 3mm
 For the usual Gaussian ensemble of random hermitian
matrices, Kazakov has
introduced a curious, but very poweful, method
\cite{Kazakov}. It consists
of adding to
the probability distribution a matrix source, which will
be set to zero at the end
of the calculation, leaving us with a simple integral
representation for
finite $N$.  However one cannot let the source go to zero
before one reaches
the final step. Let us follow the same lines for the model at
hand.
We modify the probability distribution of the matrix by a
source $A$, an
$N\times N$ hermitian matrix with eigenvalues
$(a_1,\cdots,a_N)$ :
\begin{equation}\label{AGaussian}
  P_A(C) = {1\over{Z_A}} {\exp} (-N {\rm Tr} C^{\dag}C -
N{\rm Tr} AC^{\dag}C).
\end{equation}
Next we introduce the Fourier transform of the average
resolvent with this
modified distribution:
\begin{equation}\label{U}
U_A(t) =  < {1\over N} {\rm Tr} e^{itC^{\dag}C} >
\end{equation}
from which we recover, after letting the source $A$ go to
zero,
 \begin{equation}\label{101}
  \tilde \rho(r) = \int_{-\infty}^{\infty} {dt\over{2\pi}} e^{-
itr}U_0 (t).
\end{equation}
Without loss of generality we can assume that $A$ is a
diagonal matrix. We first integrate over the
unitary matrix $V$ which diagonalizes $C^{\dag}C$ . This is
done through
the well-known Itzykson-Zuber integral over the unitary
group \cite{Itzykson}
\begin{equation}\label{IZ}
  \int dV \exp ({\rm Tr} AVBV^{\dag}) =  {\det (\exp a_i b_j)
\over
{\Delta(A) \Delta(B)}}
\end{equation}
where $\Delta(A)$ is the Van der Monde determinant
constructed with the
eigenvalues of $A$:
\begin{equation}\label{VdM}
  \Delta(A) = \prod_{i<j} ( a_i - a_j).
\end{equation}
We are then led to
\begin{eqnarray}\label{111}
U_A(t) &=& {1\over {Z_A\Delta(A)}}{1\over N}
\sum_{\alpha=1}^{N}
\int dr_1\cdots dr_N e^{it
r_{\alpha}}\Delta(r_1,\cdots,r_N)\nonumber\\
&\times &\exp
(-N\sum_{i=1}^N r_i -
N\sum_{j = 1}^{N} r_{j} a_{j}).
\end{eqnarray}
Then we have to integrate over the $r_i$'s. It is easy to
prove that
\begin{eqnarray}\label{112}
&&\int dr_1\cdots dr_N \Delta(r_1,\cdots,r_N)\exp (-
\sum_{i = 1}^{N} r_{i} b_{i}) \nonumber\\
&&=  (-1)^{N(N-1)\over {2}} (\prod_0^{N-1} {k! }
){\Delta(b_1,\cdots,b_N)
\over{(\prod_1^N {b_i})^N}}.
\end{eqnarray}
With the normalization $U_{A}(0)=1$, we could always
divide, at any
intermediate step of the calculation, the expression we
obtain for
$U_{A}(t)$ by its value at $t=0$, and thus the overall
multiplicative
factors in (3.6) and (3.7) are not needed. They are displayed
explicitly
only for the sake of completeness.

We now apply this identity to the $N$ terms of (\ref{111}),
with
\begin{equation}\label{barelation}
b_{\beta}^{(\alpha)}(t) = N ( 1 + a_{\beta} - {it \over{N}}
\delta_{\alpha,\beta})
\end{equation}
and obtain
\begin{eqnarray}\label{113}
   U_A(t) &=&{1\over{N}}\sum_{\alpha=1}^{N}
\prod_{\beta=1}^{N} ({1 + a_{\beta}\over
{1 + a_{\beta} - {it\over{N}}\delta_{\alpha,\beta}}})^N
\prod_{\beta < \gamma}{a_\beta - a_\gamma
-{it\over{N}}(\delta_{\alpha,\beta}
-\delta_{\alpha,\gamma})\over{a_\beta -
a_\gamma}}\nonumber\\
&=& {1\over{N}}\sum_{\alpha=1}^N [ {1 + a_\alpha \over {1 +
a_\alpha - {it\over{N}}}}]^N \prod_{\gamma \neq \alpha}
( {a_\alpha  - a_\gamma - {it\over N}
\over {a_\alpha - a_\gamma}})
\end{eqnarray}
This sum over $N$ terms may be conveniently replaced by
a  contour integral
in the complex plane:
\begin{equation}\label{114}
U_A(t) =  - {1\over{it}}\oint {du\over{2\pi i}}\left({1 + u
\over{1 + u -
{it\over{N}}}}\right)^N \prod_{\gamma=1}^N {u - a_\gamma
- {it\over
N}\over{ u -a_\gamma}}
\end{equation}
in which the contour encloses all the $a_\gamma$'s and no
other singularity.
  It is  now, and only now, possible to let all the
$a_\gamma$'s go to
zero. We thus obtain a simple expression for
$U_0(t)$,
\begin{equation}\label{115}
   U_0(t) = {i\over{t}}  \oint {du\over{2\pi i}}\left({1 -
{it\over{Nu}}\over
{1 - { it\over{N(1+u)}}}}\right)^N
\end{equation}
Note that this representation as a contour integral  over
one single
complex variable is  exact for any
finite $N$, including $N=1$.
\vskip 5mm
{\bf 3.2 Semi-circle law}
\vskip 3mm
  In the large $N$ limit, for finite $t$, the integrand has the
limit
$e^{{it\over{u(1-u)}}}$
and therefore for large $N$, finite $t$,  $U_0(t)$ approaches
\begin{equation}\label{116}
      U_0(t) = {1\over{it}}\int {du\over{2\pi i}}
e^{{it\over{u(1-u)}}}
\end{equation}
 By the change of variables
\begin{equation}\label{117}
     u = {1\over{2}}\left( 1 - \sqrt{1 - {4\over{z}}} \right)
\end{equation}
we have $ z = {1\over{u(1 - u)}}$.
Then the integral of (\ref{116}) becomes, after an
integration by part,
\begin{eqnarray}\label{118}
     U_0(t) &=& - \oint {dz\over{2\pi i}} u(z)
e^{itz}\nonumber\\
          &=& - {1\over{2}} \oint \left( 1 - \sqrt{1 -
{4\over{z}}}\right)
e^{itz} {dz\over{2\pi i}}\nonumber\\
   &=& \int_0^4 \sqrt{ -1 + {4\over{x}}} e^{itx}
{dx\over{2\pi}}
\end{eqnarray}
Therefore, we have from (\ref{101})
\begin{eqnarray}\label{119}
    \tilde \rho ( \lambda ) &=& \int {dt\over{2\pi}}
       e^{-it\lambda} U_0(t)\nonumber\\
     &=& {1\over{2\pi}} \sqrt{{4 - \lambda \over{\lambda}}}
\end{eqnarray}
which leads to the expected semi-circle law
\begin{equation}
     \rho(\mu) =  |\mu | \tilde \rho( \mu^2) =
{1\over{2\pi}}\sqrt{ 4 - \mu^2}
\end{equation}
\vskip 5mm
{\bf 3.3 Behavior near the origin}
\vskip 3mm
For $\lambda$ small however, we need to control the large
$N$ limit of
$U_0(t)$, for large $t$.
In this regime the simplified form (\ref{116}) of (\ref{115})
is not valid.
This is why the semi-circle law is  modified at the origin.
Keeping in mind that we shall consider $\lambda$'s of order
$1/N$, we write
\begin{eqnarray}
    \tilde\rho(\lambda^2) &=& N^2 \int_{-\infty}^{+\infty}
{d\tau \over{2\pi}}
e^{-i\tau N^2 \lambda^2}U_0(N^2 \tau)\nonumber\\
      &=& {N\over{i}} \int_{- \infty}^{+\infty}
       {d\tau\over{2\pi}} e^{-i\tau x^2}{1\over{\tau}}
        \oint
        {du\over{2\pi i}}\left( { 1 - {1\over{Nu}}
\over{1 - {1\over{N(u+i\tau)}}}}
        \right)^N
\end{eqnarray}
in which we have defined
\begin{equation}
x = N \lambda
\end{equation}
In the large $N$, small $\lambda$ but finite $x$, limit, we
thus obtain
\begin{equation}
      \tilde\rho(\lambda^2) = {N\over{i}}\int_{-\infty}^{\infty}
      {d\tau\over{2\pi}} \oint {du\over{2\pi i}} {1\over{\tau}}
e^{-i\tau x^2 +
{1\over{i\tau + u}} - {1\over{u}}}.
\end{equation}
If we calculate instead
\begin{equation}
I(x^2) = {\partial  \tilde\rho\over{\partial x^2}}
\label{process1}
\end{equation}
we find, after changing $u$ to $iu$ and then $\tau$ to $\tau -
u$,
\begin{equation}
     I(x^2) = - Ni \int_{-\infty}^{+\infty} {d\tau\over{2\pi}}
      e^{-{i\over{\tau}} - i \tau x^2 } \oint {du\over{2\pi i}}
 e^{{i\over u} + iux^2}
\label{process2}
\end{equation}
Using the standard definitions of Bessel functions, we have
\begin{equation}
   \int_{-\infty}^{+\infty} {d\tau\over{2\pi}} e^{-ix(\tau +
{1\over \tau})}
    = - {1\over{2}}J_1(2x)
\end{equation}
\begin{equation}
   \oint {du\over {2 i\pi}} e^{ix (u+{1\over u})} = i J_1 (2x)
\end{equation}
and thus
\begin{eqnarray}
     I(x^2) &=& - {N\over {2 x^2}} J_1^2(2x)
\nonumber\\
             &=& {N\over{2}} {d\over{dx^2}} [ J_0^2(2x) +
J_1^2(2x) ]
\end{eqnarray}
Therefore the integration with respect to $x^2$ is
immediate, and we find
\begin{equation}\label{crossover}
   \rho(\lambda) = {N |\lambda|\over{2}} [ J_0^2(2N\lambda)
+ J_1^2(2N\lambda)]
\end{equation}
in the cross-over regime in which $N\lambda$ is finite. The
oscillatory
behavior we expected is described by Bessel functions
\vskip 5mm
{\bf 3.4 The edge of the semi-circle}
\vskip 3mm
 It is easy to apply this same method for studying the
cross-over at the
other edge of the distribution, in the vicinity of the end
point $\mu =2$
of the semi-circle.
The derivative of the density of state
 $\tilde\rho(\mu^2)$ with respect to $\mu^2$ is given by
\begin{equation}\label{rho1}
   {\partial  \tilde\rho(\mu^2)\over{\partial \mu^2}}
 =  \int_{-\infty}^{+\infty} {dt\over{2 \pi}} \oint {du\over{2
\pi i}}
    \left[{1 - {it\over{Nu}}\over{1 - {it\over{N(1 +
u)}}}}\right]^N
e^{-it\mu^2 }
\end{equation}
Changing  $t$ to $Nt$, and then $t$ to $t-iu$, and also $u$ to
$-iu$,
we obtain the factorized expression,
\begin{equation}\label{rho prime}
   {\partial  \tilde\rho(\mu^2) \over{\partial \mu^2}} =
   - i N \int_{-\infty}^{+\infty} {dt\over{2 \pi}}
({t\over{1 - it}})^N e^{-iN\mu^2 t} \oint {du\over{2 \pi i}} ({1
- iu \over
{u}})^N e^{ iN\mu^2 u}
\end{equation}
The integration over $t$ is easily expressible with the help of
a Laguerre
polynomial:
\begin{eqnarray}\label{IN1}
     I_N(\mu^2) &=&   \int_{-\infty}^{+\infty} dt
({t\over{1 - it}})^N e^{-iN\mu^2 t}\nonumber\\
     &=& - 2\pi i^N e^{-N\mu^2} L_N^\prime(N\mu^2)
\end{eqnarray}
The contour integral over $u$ turns out to be also
expressible as a
derivative of a Laguerre polynomial;
we end up with the  simple expression for (\ref{rho1}),
\begin{equation}\label{rho2}
   {\partial \tilde\rho (\mu^2)\over{\partial \mu^2}} = - N
e^{-N\mu^2}
[L_N^\prime(N\mu^2)]^2
\end{equation}

Using standard identities for Laguerre polynomials it is easy
to verify
that this leads to
\begin{equation}
    \tilde\rho (\mu^2) = N e^{-N \mu^2} [ L_N(N\mu^2)L_{N-
1}^\prime(N\mu^2)
- L_{N-1}(N\mu^2) L_N^\prime(N\mu^2)]
\end{equation}
The orthogonal polynomial method , which led to the
expression  (2.8),
may be cast into this form through Christoffel-Darboux
identity. However
for our purpose, the cross-over distribution near the edge
of the
semi-circle, it is much more convenient to return to
 the integral $I_N$ of (\ref{IN1}) and to use  the saddle point
method:
\begin{equation}
    I_N = \int_{-\infty}^{+\infty} dt e^{-NS_{eff}}
\end{equation}
where $S_{eff}$ is given by
\begin{equation}
    S_{eff} = - {\ln t} + {\ln (1 - it)} + i\mu^2 t .
\end{equation}  The
saddle points $ t_c$  become degenerate at $\mu = 2$, since
\begin{equation}
   t_c = {-i \pm \sqrt{ {4\over{\mu^2}} - 1}\over{2}}.
\end{equation}
Then we  change  variables to
\begin{eqnarray}
    \mu &=&  2 + N^{-\alpha} x,\nonumber\\
    t &=& - {i\over{2}} + N^{-\beta} \tau
\end{eqnarray}
and expand $S_{eff}$ up to $\tau^3$. This  leads to
\begin{eqnarray}\label{taus}
S_{eff}(t) &=& 2 +{\pi i\over{2}} + 2N^{-\alpha} x
\nonumber\\
&+&
i {16\over{3}} \tau^3 N^{-3\beta}
   + 4i N^{-\alpha - \beta} \tau x
\end{eqnarray}
We thus find that there is a large $N$, finite $x$ limit,
provided we
fix the two unknown parameters $\alpha$ and $\beta$ to
\begin{equation}
   \alpha = {2\over{3}}, {\hskip 5mm} \beta = {1\over{3}}
\end{equation}
We repeat this for the $u$-integral of (\ref{rho prime}).  We
then find that
the leading terms of (\ref{taus}) of order 1, as well as the
term   $2x
N^{-2/3}$,  cancel  with terms of opposite
signs  in the $u$-integral.
Thus we obtain the following expression for the density of
state near the
critical
value $\mu = 2$,
\begin{equation}\label{edge Airy}
    {\partial \tilde \rho(\mu^2)\over{\partial \mu^2}}
= - N^{1\over{3}}4^{-
{4\over{3}}}|A_i[4^{1\over{3}}N^{2\over{3}}(\mu - 2)]|^2
\end{equation}
where the Airy function $A_i(z)$ is defined as
\begin{equation}
     A_i[ (3a)^{-1/3} x] = {(3a)^{1/3}\over{\pi}}
\int_0^\infty \cos (at^3 + xt)dt.
\end{equation}
This Airy function is  smoothly decreasing for $\mu > 2$ but
it gives
oscillations
for $\mu < 2$.

\vskip 5mm
{\bf 3.5 Two-point correlation}
\vskip 3mm
The application of this method to the two-point correlation
function
is straightforward; it leads to a  compact and useful integral
representation.
Let us briefly describe
the procedure.
Introducing the same source $A$ in the probability
distribution as in
(\ref{AGaussian}), we consider the Fourier transform of the
average two-point
correlation:
\begin{equation}\label{U100}
   U_A^{(2)}(t_1,t_2) = < {1\over{N}} {\rm Tr} e^{it_1 C^{\dag}
C}
  {1\over{N}}{\rm Tr} e^{it_2 C^{\dag} C} >
\end{equation}
from which we shall compute the two-point correlation
function
$\tilde\rho^{(2)}(\lambda_1,\lambda_2)$,
after letting the source $A$ goes to zero.
The normalization conditions are then
\begin{eqnarray}\label{U101}
    &&U_A^{(2)}(t_1,t_2) = U_A^{(2)}(t_2,t_1),\nonumber\\
    &&U_A^{(2)}(t,0) = U_A^{(1)}(t),\nonumber\\
    &&U_A^{(1)}(0) = 1.
\end{eqnarray}
After performing the Itzykson-Zuber integral over the
unitary group as in
(\ref{111}),
 we obtain through the same procedure,
\begin{equation}\label{U102}
   U_A^{(2)}(t_1,t_2) =
{1\over{N^2}}\sum_{\alpha_1,\alpha_2} \int
\prod_{i=1}^N d r_i {\Delta(r)\over{\Delta(A)}}e^{-
N\sum_{i=1}^N
 r_i (1 + a_i)
 + it_1 r_{\alpha_1} + it_2 r_{\alpha_2}}
\end{equation}
where we omitted the overall normalization, as we are
allowed to do (as
exoplained earlier).
The integration of $r_i$ is again done with the use of
(\ref{U103}), in
which now
\begin{equation}\label{U103}
   b_\beta^{(\alpha_1,\alpha_2)} = N(1 + a_\beta -
{1\over{N}}(it_1
\delta_{\beta,
\alpha_1} + it_2 \delta_{\beta, \alpha_2})).
\end{equation}
Thus we have the following expression for
$U_A^{(2)}(t_1,t_2)$ after
restoring the normalization,
\begin{eqnarray}\label{U104}
  U_A^{(2)}(t_1,t_2)
&=& {1\over{N^2}}\sum_{\alpha_1,\alpha_2 = 1}^N
 \prod_{\beta = 1}^N \left[{1 + a_\beta\over{1 + a_\beta -
({it_1\over{N}}
\delta_{\beta,\alpha_1} + {it_2\over{N}}
\delta_{\beta,\alpha_2})}}\right]^N
\nonumber\\
&\times& \prod_{\beta<\gamma}{ a_\beta -
a_\gamma - {it_1\over{N}}(\delta_{\beta \alpha_1} -
\delta_{\gamma \alpha_1})
- {it_2\over{N}}(\delta_{\beta \alpha_2} - \delta_{\gamma
\alpha_2})
\over{(a_\beta - a_\gamma)}}\nonumber\\
\end{eqnarray}
Keeping track of all the terms  in which the Kronecker
$\delta_{\alpha,\beta}$'s do not  vanish,
 we obtain
\begin{eqnarray}\label{U105}
  &&U_A^{(2)}(t_1,t_2) = {1\over{N}}U_A^{(1)}(t_1 + t_2) +
{1\over{N^2}} \sum_{\alpha_1 \neq \alpha_2}^N\left[{(1 +
a_{\alpha_1})
(1 + a_{\alpha_2})\over{(1 + a_{\alpha_1} - {it_1\over{N}})(1
+ a_{\alpha_2}
- {it_2\over{N}})}}\right]^N\nonumber\\
&\times& {(a_{\alpha_1} - a_{\alpha_2} - {it_1 -
it_2\over{N}})\over
{(a_{\alpha_1} - a_{\alpha_2})}}\prod_{\gamma\neq
(\alpha_1,\alpha_2)}^N
{(a_{\alpha_1} - a_{\gamma} - {it_1\over{N}})(a_{\alpha_2} -
a_{\gamma}
-{it_2\over{N}})\over{(a_{\alpha_1} -
a_{\gamma})(a_{\alpha_2} - a_{\gamma})}}
\end{eqnarray}
The Fourier transform of
the first term of (\ref{U105}) corresponds to
\begin{equation}
   {1\over{N(2\pi)^2}}\int dt_1 dt_2 e^{-it_1 \lambda_1 -
it_2 \lambda_2}
U_0^{(1)}(t_1 + t_2) = {1\over{N}}
\delta(\lambda_1 - \lambda_2) \rho_0(\lambda_1)
\end{equation}
It could thus be omitted for  $\lambda_1 \neq \lambda_2$
but, remarkably
enough, the contour integral that we shall now consider,
will be simpler if
we retain this term. Indeed let us consider the integral over
two complex
variables $u$ and $v$
\begin{eqnarray}
&& U_A^{(2)}(t_1,t_2) = {1\over{ t_1 t_2}}\oint {du
dv\over{(2\pi i)^2}}
\left[{(1+u)(1+v)\over{(1+u-{it_1\over{N}})(1+v-
{it_2\over{N}})}}\right]^N
\nonumber\\
&&\prod_{\gamma} \left( {u - a_\gamma -
{it_1\over{N}}\over
{u - a_\gamma}}\right)
 \left( {v - a_\gamma - {it_2\over{N}}\over{v -
a_\gamma}}\right)
{(u - v - {it_1 - it_2 \over{N}})( u - v) \over{( u - v -
{it_1\over{N}})
( u - v + {it_2 \over{N}})}} .
\end{eqnarray}
It is straightforward to verify that this expression reduces
exactly to
(\ref{U105}), provided we choose the following contours: we
first integrate
over a contour in $v$ which circles around the
$a_{\alpha}$'s and no other
singularity. Then taking a residue at  $ v=a_{\alpha_2}$ , we
integrate
over a contour in $u$ which surrounds the $(N-1)$ poles
$u = a_{\alpha_1}$, with  ${\alpha_1\neq \alpha_2}$; these
poles yield the
sum over  ${\alpha_1}$ and ${\alpha_2} $ of (\ref{U105}).
The contour in $u$
has to surround also  the pole $ u= a_{\alpha_2} -{it_2
\over{N}}$.
Remarkably, this last pole reproduces exactly the first term
${1\over{N}}U_A^{(1)}(t_1 + t_2) $ of  (\ref{U105}).

We are now again in position to  let all the $a_\gamma$'s  go
to zero:
\begin{eqnarray}\label{U106}
U_0^{(2)}(t_1,t_2) &=& {1\over{t_1 t_2}}\oint {du
dv\over{(2\pi i)^2}}
\left[{(1 + u)(1 + v)\over{
(1 + u - {it\over{N}})(1 + v - {it_2\over{N}})}}\right]^N
( 1 - {it_1\over{Nu}})^N
\nonumber\\
&\times& ( 1 - {it_2\over{Nv}})^N
 \left[ 1 - {t_1 t_2 \over{N^2(u - v - {it_1\over{N}})
( u - v + {it_2\over{N}})}} \right]
\end{eqnarray}
The last bracket in (\ref{U106}) is a difference of two terms.
Keeping the
one in this bracket  we obtain the
 disconnected part $U_0^{(1)}(t_1)U_0^{(1)}(t_2))$ of
$U_0^{(2)}(t_1,t_2)$.
The second term gives  thus the connected two-point
correlation
$\tilde\rho_{c}^{(2)}(\lambda_1,\lambda_2)$.

 In the large $N$ limit, the connected two-point correlation
may be then
immediately
obtained. For finite $t_1$ and $t_2$  we have, in the large
$N$ limit
\begin{equation}\label{U107}
 U_c^{(2)}(t_1,t_2) = - {1\over{N (2\pi i)^2}}\oint {du
dv\over{(u - v)^2}}
   e^{-{it_1\over{u(1 + u)}} - {it_2\over{v(1 + v)}}}
\end{equation}
Changing  variables to
\begin{equation}\label{U108}
z_1 = {1\over{u(1 + u)}}, {\hskip 5mm} z_2 = {1\over{v(1 +
v)}}
\end{equation}
and denoting
\begin{equation}\label{U109}
  u(z) = {-1 + \sqrt{ 1 + {4\over{z}}}\over{2}},
\end{equation}
we have
\begin{equation}\label{U110}
  U_c^{(2)}(t_1,t_2) = {t_1 t_2\over{N^2}} \int dz_1 dz_2
e^{-it_1 z_1 - it_2 z_2} \ln [ u(z_1) - u(z_2) ].
\end{equation}
Indeed we have used ${\partial \over{\partial z_2}}
{\partial \over{\partial z_1}}
\ln ( u - v) = {1\over{(u - v)^2}} ({\partial u\over{\partial
z_1}})
({\partial v \over{\partial z_2}})$, and then integrated by
part over $z_1$
and $z_2$.
We then  Fourier transform over $t_1,t_2$:
\begin{equation}\label{U111}
\tilde\rho_{c}^{(2)}(\lambda_1,\lambda_2) = - {1\over{N^2}}
{\partial^2\over{\partial \lambda_1 \partial \lambda_2}}\ln
[ u(\lambda_1)
- u(\lambda_2)]
\end{equation}
This result could be derived by other methods, and indeed has been
obtained
in a somewhat different form by Ambjorn and Makeenko and others \cite{amb}.
The derivation given here however gives the
correlation function even for finite $N$.
One can check
easily that the derivation that we have given here through Kazakov's
representation, could
be repeated for calculating the two-point correlation
function in the
unitary ensemble, in which the universality of the two-
point correlation
function has been studied
\cite{ambjorn, BZ1,bz2,been,eyn,forr,bhz}.

It is not difficult to verify that the above expression for the
connected
two-point correlation is a compact version of the one that
we would have
deduced from the orthogonal polynomial method, with the
kernel
$K(\lambda_1, \lambda_2)$ in (\ref{C4}). Indeed, after
letting the
$a_\gamma$'s go to zero, we determine the
connected two-point correlation function to be
\begin{eqnarray}\label{U112}
   \tilde\rho_{c}^{(2)}(\lambda_1,\lambda_2) &=&
- \int {dt_1 dt_2\over{(2\pi)^2}} \oint {du dv\over{(2\pi
i)^2}}
 e^{-iNt_1 \lambda_1 - iNt_2 \lambda_2}
\nonumber\\
&& ({u - it_1\over{1 + u -it_1}})^N ( {v - it_2\over{1 + v -
it_2}})^N
( {1 + u\over{u}})^N ( {1 + v\over{v}})^N
\nonumber\\
&\times & {1\over{(u - v - it_1)(u - v + it_2)}}
\end{eqnarray}
This expression takes  a factorized form if we shift $t_1$ to
$t_1 - iu$ and $t_2$ to $t_2 - iv$:
\begin{equation}
    \tilde\rho_{c}^{(2)}(\lambda_1, \lambda_2) = -
I(N\lambda_1,
N\lambda_2) I(N\lambda_2, N\lambda_1)
\end{equation}
in which we have defined
\begin{equation}
I(\lambda_1,\lambda_2) = \int {dt\over{2\pi}} \oint
{dv\over{2\pi i}} \left( {t\over{t + i}}\right)^N
\left( {1 - iv\over{v}}\right)^N {1 \over {t - v}}
e^{-it\lambda_1 + iv\lambda_2}
\end{equation}
Taking the residues at the poles $t= -i$ and $v = 0$,
we find
 \begin{equation}
     I(\lambda_1, \lambda_2) = e^{-
\lambda_1}\sum_{n=0}^{N-1}
L_n(\lambda_1) L_n(\lambda_2)
\end{equation}
Thus $I(\lambda_1, \lambda_2)$ is simply an integral
representation for the kernel $K(\lambda_1, \lambda_2)$ in
(\ref{C4}).
The connected two-point correlation function
$\rho_c^{(2)}(\mu,\nu)$ is
expressible through the Christoffel-Darboux identity,
\begin{eqnarray}
    \rho_c^{(2)}(\mu,\nu) &=& \mu \nu \tilde
\rho_c^{(2)}(\mu^2, \nu^2)
\nonumber\\
&=& \mu \nu {[L_N(N\mu^2) L_{N-1}(N\nu^2) -
L_N(N\nu^2) L_{N-1}(N\mu^2)]
\over {(\mu^2 - \nu^2)^2}}
\end{eqnarray}
\sect{Use of Grassmannian variables}

In the previous section we have used a source
representation, which is
powerful and simple. However we would like to investigate
the same question
of the edge behavior near the origin for more complicated
ensembles of
block matrices which arise when the randomness is due to
random couplings
between neighbors on a lattice.  Unfortunately we have not
found any simple
extension of Kazakov's method and we have to use
Grassmannian variables, as
often in disordered systems, to solve the problem
\cite{Brgrass}.
 Before going to a
lattice of matrices we  return to the case that we have
solved in the previous
section, in order to show how to recover the same results
from this new
method. We shall then extend it to lattices in the next
section.

  We begin with the identity
\begin{eqnarray}\label{A1}
   {M_{ab}^{-1}} &=& - i N \int
\prod_{1}^{N}(du_cdu_c^{*}dv_{c}dv_{c}^{*})\nonumber\\
&\times & u^{*}_{a}u_{b} {\exp} [ i \sum_{c,d}
N(u^{*}_{c}M_{cd}u_{d}+ v^{*}_{c}M_{cd} v_{d})]
\end{eqnarray}
in which the $u$'s are commuting variables and the $v$'s
are
Grassmannian.
Indeed with the normalization
\begin{equation}\label{A2}
   \int dv dv^{*} v v^{*} = {1\over{\pi}}
\end{equation}
one verifies that
\begin{equation}
     \int du du^{*}dvdv^{*} {\exp} [-(u^{*}u + v^{*}v)] = 1
\end{equation}
In order to apply this to the matrix
\begin{equation}\label{twoBYtwo}
   ( z - M ) =\left(\matrix{z& -C^{\dag}\cr
                 -C& z \cr}\right),
\end{equation}
which is $2N\times 2N$, we have to decompose the $2N$-
component
vectors $u$ and
$v$, into $N$-component vectors;
\begin{equation}\label{AA2}
    u = \left(\matrix{a\cr b}\right), v =\left(\matrix{\alpha\cr
\beta}\right)
\end{equation}
and we obtain
\begin{eqnarray}\label{oneG}
   & &{1\over{2N}} {\rm Tr} {1\over{z - M}}  = - {i\over 2}
\int
\prod_{1}^{N}
(da_cda^{*}_cdb_cdb^{*}_cd\alpha_{c}d\alpha^{*}_cd\beta_{c}d
\beta
^{*}_c)
\nonumber\\
& & [(a^{*}\cdot a) + (b^{*}\cdot b)]
   {\exp}  ( iNz[(a^{*}\cdot a)+(b^{*}\cdot b)+
(\alpha^{*}\cdot \alpha) + (\beta^{*}\cdot
\beta)]\nonumber\\
& &-iN[a^{*}_cC^{\dag}_{cd}b_d
+\alpha^*_cC^{\dag}_{cd}\beta_d +
b^*_cC_{cd}a_d + \beta^*_cC_{cd}\alpha_{d}])
\end{eqnarray}
The average over the Gaussian distribution for the matrix
$C$
\begin{equation}\label{PGaussian}
  P(C) = {1\over{Z}} {\exp} (-N {\rm Tr} C^{\dag}C)
\end{equation}
is then easily performed, since

\begin{equation}
< {\exp} i N {\rm Tr}(\lambda C + \mu C^{\dag})> = {\exp}(-N
{\rm
Tr} (\lambda \mu))
\end{equation}
This gives for the average resolvent
\begin{eqnarray}\label{A4}
  G(z) &=& < {1\over{2N}} {\rm Tr}{1\over{z - M}}
>\nonumber\\
     &&= -{i\over{2}}\int
\prod_{1}^{N}(da_{c}da^{*}_cdb_cdb^{*}_cd\alpha_{c}
d\alpha^{*}_cd\beta_cd\beta^{*}_c)
[(a^{*}\cdot a)+(b^{*}\cdot b)]\nonumber\\
&& {\exp}( iNz[(a^{*}\cdot a) + (b^{*}\cdot b)+(\alpha^{*}\cdot
\alpha) +
(\beta^{*}\cdot \beta)]
- N[(a^{*}\cdot a)(b^{*}\cdot b)\nonumber\\
&& - (\alpha^{*}\cdot \alpha)(\beta^{*}\cdot \beta) +
(\alpha^{*}\cdot a)(b^{*}
\cdot \beta)
+ (\beta^{*}\cdot b)(a^{*}\cdot \alpha)])
\end{eqnarray}

Note that there is a minus sign in front of the four Fermi
interaction
due to the Grassmannian algebra.
This four Fermi term may be replaced by an additional
integration,
since
\begin{equation}\label{sigma}
{\exp} N(\alpha^{*}\cdot \alpha)(\beta^{*}\cdot \beta) =
{N\over{\pi}}\int
d^{2}\sigma {\exp}\left ( -N[\sigma^{*}\sigma +
\sigma^{*}(\alpha^{*}\cdot \alpha) + \sigma (\beta^{*}\cdot
\beta)]\right )
\end{equation}
Substituting this into (\ref{A4}) we can now perform
the integration over the anti-commuting variables,
\begin{eqnarray}
&&\int \prod_{c = 1}^{N}
(d\alpha_cd\alpha^*_cd\beta_cd\beta^*_c)
{\exp}(-N[(\alpha^*\cdot \alpha)(\sigma^*-iz)
 + (\beta^*\cdot \beta)(\sigma - iz)]\nonumber\\
&&-N[(\alpha^*\cdot a)(b^*\cdot \beta)+(\beta^*\cdot b
)(a^*\cdot \alpha)])\nonumber\\
&&=({N\over{\pi}})^{2N} det \left( \matrix{\sigma^*-iz&
|a><b|\cr
|b><a|& \sigma - iz}\right)\nonumber\\
&& = ({N\over{\pi}})^{2N}[(\sigma^*-iz)(\sigma-iz)]^{N-1}
[(\sigma^*-iz)(\sigma-iz)
-(a^*\cdot a)(b^*\cdot b)]\nonumber\\
\end{eqnarray}
We are then led to
\begin{eqnarray}
 &&G(z) = - {iN\over{2\pi}}({N\over{\pi}})^{2N}\int
d^2\sigma
\prod_1^N
(da_cda^*_cdb_cdb^*_c)[(a^*\cdot a)+(b^*\cdot
b)]\nonumber\\
&&[(\sigma^*-iz)(\sigma-iz)]^{N-1}[(\sigma^*-iz)(\sigma-iz)-
(a^*\cdot a)
(b^*\cdot b)]
\nonumber\\
&&{\exp}(iNz[(a^*\cdot a) + (b^*\cdot b)]-N[(a^*\cdot
a)(b^*\cdot b) +
\sigma^*\sigma])
\end{eqnarray}
The integrand is a function of the lengths of the complex
vectors
$a$ and $b$;
integrating over the angles:
\begin{equation}
\int \prod_1^N(da_cda^*_c)f(a^*,a)= {\pi^N\over{(N-
1)!}}\int_{0}^{\infty} dx
x^{N-1}f(x)
\end{equation}\label{x1}
we end up with
\begin{eqnarray}\label{A8}
G(z) &=& -{i\over{2\pi}} {N^{2N+3}\over{(N!)^2}}\int
d^2\sigma
\int_{0}
^{\infty} dx \int_{0}^{\infty} dy (xy)^{N-1}(x+y)
\nonumber\\
&&[(\sigma^*-iz)(\sigma-iz)]^{N-1}
[(\sigma^*-iz)(\sigma-iz)-xy]\nonumber\\
&&{\exp}(N[iz(x+y)-xy-
\sigma^*\sigma])
\end{eqnarray}
This representation of the average resolvent in terms of an
integral
over four variables is exact for any $N$.

Let us note that, if we had calculated, instead of the average
resolvent, the average value of 1, we would have obtained
by the
same method, the identity:
\begin{eqnarray}\label{A9}
 &&{1\over{\pi}}{N^{2N+3}\over{(N!)^2}}\int d^2\sigma
\int_0^\infty dx
\int_0^\infty dy (xy)^{N-1}
[(\sigma^*-iz)(\sigma-iz)]^{N-1}\nonumber\\
&&[(\sigma^*-iz)(\sigma-iz)-xy]{\exp}(N[iz(x+y)-xy-
\sigma^*\sigma]) = 1
\end{eqnarray}
This identity should manifestly hold, but it is not quite
trivial
to check it; since it provides an interesting verification of
the
consistency
of the formation, let us note that one can derive it by
replacing, in
the l.h.s.
of ({\ref{A9}}), the bracket $[(\sigma^*-iz)(\sigma-iz)-xy]$
by
\begin{equation}
  {1\over{N^2}}({\partial^2\over{\partial \sigma \partial
\sigma^*}}-
{\partial^2\over{\partial x\partial
y}})+{iz\over{N}}({\partial\over{\partial \sigma}}
+{\partial\over{\partial \sigma^*}}+{\partial\over{\partial
x}}+{\partial\over{
\partial y}})\nonumber
\end{equation}
followed by integrations by part and a few lengthy
manipulations.
This identity will be useful in a few moments.

The expression ({\ref{A8}}) of $G(z)$ is well suited to study
the
large $N$-limit.
The integrand involves a factor $\exp (-NS)$, with
\begin{equation}\label{saddle}
S(\sigma, \sigma^*,x,y) = -iz(x+y) + xy + \sigma^*\sigma -
{\ln}[xy(\sigma^*-iz)(\sigma-iz)]
\end{equation}
and the large $N$ limit is therefore governed by
the saddle-point at which $S$ is stationary.
The equations for the saddle-point lead, away from the
vicinity of
$z=0$, to
the equations
\begin{equation}\label{largeN1}
x_c=y_c=\sigma_c=\sigma_c^*={1\over{2}}[iz+\sqrt{4-z^2}]
\end{equation}
The sign of the square root is chosen so that the imaginary
part of
$G(z)$ above
the cut is negative; note that in the saddle-point method
$\sigma$
and $\sigma^*$ become independent complex variables.
In order to obtain $G(z)$ with the proper normalization we
should
include the
Gaussian fluctuations around this saddle-point.
It presents no difficulty; however we can bypass the whole
calculation,
if we note that the integrand in ({\ref{A8}) differs from that
of
(\ref{A9}) by a factor $-i(x+y)/2$; if we had calculated
(\ref{A9})
by the same saddle-point technique we would have
obtained one,
a very cumbersome method to get one indeed. But this
immediately
tells us that, for $N$ large,
\begin{equation}
  G(z) = {-{i\over{2}}}(x_c+y_c) = {1\over{2}}[z - i\sqrt{4-
z^2}]
\end{equation}
from which one recovers the semi-circle law
\begin{equation}\label{density}
    \rho(\lambda) = -{1\over{\pi}} {\rm Im} G(\lambda + i0) =
{1\over{2\pi}}\sqrt{4-\lambda^2}
\end{equation}
However this solution is not valid near the vicinity of the
origin.
Indeed
the saddle-point equations gives $z(x_c-y_c)=z(\sigma_c-
\sigma^*_c)=0$;
another way of realizing that there is a problem near $z=0$,
is to
calculate the determinant of the Gaussian fluctuations near
the
saddle-point, which vanishes when $x_c^4=1$, i.e. for
$z=\pm 2$ or
$z=0$.
Near the end-points of the semi-circle the phenomenon is
well-known and leads to an Airy type cross-over function
in a
region of size $N^{-2/3}$ near the edge, which we have
discussed in the
previous section.
Near the origin, we need a separate analysis.
We will focus on a range of size $1/N$, in which the variable
\begin{equation}
    \zeta=Nz
\end{equation}
is finite.
We write $\sigma = u+iv$, $\sigma^*=u-iv$ and translate the
real $u$-
contour
by $iz$; this gives
\begin{eqnarray}\label{exact}
&&G(\zeta) = -{i\over{2\pi}}{N^{2N+3}\over{(N!)^2}}\int du
dv
\int_{0}^\infty
dx \int_{0}^{\infty}dy (xy)^{N-1}(x+y)[u^2+v^2]^{N-
1}\nonumber\\
&&[u^2+v^2-xy]{\exp}(-N[xy+u^2+v^2]){\exp}(i\zeta(x+y))
{\exp}(-2iu\zeta + {1\over{N}}\zeta^2)\nonumber\\
\end{eqnarray}
Going into radial variables for $u$ and $v$, and changing
$x$, $y$ to
$p$ and
$q$ with $xy=p$, $x+y=2\sqrt{p}q$ (Jacobian $J=(q^2-1)^{-
1/2}$,
domain
of integration $p>0$, $q>1$), we obtain easily, if we drop
terms of
order $1/N$,
\begin{eqnarray}
   &&G(\zeta) = - {iN^2\over{4\pi^2}}\int_{0}^{2\pi} {d\theta}
\int_{1}^{\infty} dq {q\over{\sqrt{q^2-1}}}\int_{0}^\infty
{dr\over{r}}
\int_{0}^{\infty}{dp\over{\sqrt{p}}}(r-p)\nonumber\\
&&{\exp}(-N[p+r-2-\ln(pr)]){\exp}(2i\zeta q \sqrt{p})
{\exp}
[-2i\zeta \sqrt{r} {\cos \theta}]
\end{eqnarray}
The remaining integrals over $p$ and $r$ are governed, in
the large
$N$ limit,
by their saddle-points at $p=r=1$. However the presence of
the odd
factor
$(r-p)$ in the integrand implies to expand beyond the
leading non-Gaussian order. Changing variable $q$ to $\cosh
\phi$, we
integrate $\phi$
and $\theta$,
\begin{eqnarray}\label{K1}
    \int_{0}^{\infty} d{\phi} {\cosh \phi} {e}^{2i\zeta \sqrt{p}
 {\cosh}\phi}
   &=& K_1(-2i\zeta\sqrt{p})\nonumber\\
   &=& -{\pi\over{2}} [ J_1 (2 N \mu\sqrt{p}) + i N_{1} (2 N
\mu\sqrt{p})]
\end{eqnarray}
in which $N \mu$ is the real part for $\zeta$, i.e. the
eigenvalue of
the matrix multiplied by $N$;
\begin{equation}
     \int_{0}^{2\pi} {\exp}(-2i N \mu \sqrt{r} {\cos
\theta})d\theta =
     2\pi J_{0}(2 N \mu \sqrt{r})
\end{equation}
we have for the imaginary part of $G(\zeta)$
\begin{eqnarray}\label{J1J0}
  \rho(\mu)&=& -{N^2\over{4\pi}}\int_{0}^\infty
{dr\over{r}}\int_{0}^\infty
{dp\over{\sqrt{p}}}(r-p)\nonumber\\
&& {\exp}(-N[p+r-2 -{\ln}(pr)])J_{1}(2 N \mu \sqrt{p})
J_{0}(2 N \mu \sqrt{r})\nonumber\\
&&
\end{eqnarray}
Expanding the Bessel functions $J_1(2 N \mu \sqrt{p})$ and
$J_{0}(2 N \mu \sqrt{r})$,
with $r=1 + r'$ and $p=1 + p'$ up to  linear order, we get
\begin{equation}
   J_{1}(2 N \mu \sqrt{p}) = J_1(2 N \mu) + p'[ N \mu J_0(2 N
\mu) -
{1\over{2}}J_1(2 N \mu)]
\end{equation}
\begin{equation}
   J_{0}(2 N \mu \sqrt{r}) = J_0(2 N \mu) - N \mu r' J_1(2 N
\mu)
\end{equation}
Then, by the Gaussian integration over $p'^2$ and $r'^2$,
we have
\begin{equation}\label{osdens}
\rho(\mu) = {N |\mu |\over{2}}[J_0^2(2 N \mu) + J_1^2(2 N
\mu)]
\end{equation}
Since $J_0(x)\simeq 1$ and $J_1(x) \simeq x/2$ near $x = 0$,
the
density of
state $\rho(\mu)$ is proportional to $\mu$ for small $\mu$.
 The  density of states that we have found agrees with
(\ref{crossover}).
 If we
average over these oscillations with the appropriate width,
we find
$<\rho(\mu)> = 1/\pi$, the value that it takes at the origin in
the
saddle-point method for the large $N$ limit (\ref{density}).


\sect{Ring of matrices}
 We now extend the previous Grassmannian method to the
case of a lattice of
matrices. We consider a ring of $L$ points, with $L$ even, in
which the
neighboring sites are coupled by  complex $ N \times N$
matrices. The
previous section corresponded to $L=2$, a lattice of two
points, coupled by
the matrix $C$ for one orientation of the link, and $C^{\dag}$
for the
opposite orientation. The simplest extension, $L=4$  consists
of a $4 N
\times 4 N$ matrix
given by
\begin{equation}\label{B1}
    M = \left(\matrix{0&C_1^{\dag}&0&C_4\cr
         C_1&0&C_2^{\dag}&0\cr
         0&C_2&0&C_3^{\dag}\cr
         C_4^{\dag}&0&C_3&0\cr}\right)
\end{equation}
where $C_i$ is a $N\times N$ complex matrix. This matrix $M$
represents a
random hopping between the nearest neighbour sites of a
lattice of four
points on a
ring.
The disorder is off-diagonal, and the hopping terms are
  $N\times N$ random complex matrices. This matrix $M$ has
also pairs of opposite real eigenvalues, since it anticommutes
with the
``$\gamma_5$"
matrix. However this $L=4$ case
 ({\ref{B1}) happens to
be reducible to
the previous case  through the orthogonal transformation
which exchanges
indices two and three; namely
if we  change $M$ to $P^{-1}M P$ with
$P(1,1)=P(2,3)=P(3,2)=P(4,4)=1$, and all other $P(i,j)$'s
equal zero, one
sees
easily that the problem is mapped into the $L=2$ case with
$N$ replaced by
$2N$.  This is not true however for larger rings.

For $L$ even the matrix $M$
has again pairs of opposite real eigenvalues since  it anti-
commutes with
the $``\gamma_5$" matrix made of $L$ block matrices
consisting successively
of the unit matrix and of minus the unit matrix. One can
thus consider
again the problem of the behaviour of the density of
eigenvalues in the
scaling range near the origin. We shall prove that the
previous result
still holds up to scale factors.

The formulation of the previous section
for $ 2N \times 2N$ matrices, may be easily extended
to the $LN \times LN$ matrices, corresponding to a   one
dimensional lattice.
The $2N$ component vectors $u$ and $v$ in (\ref{AA2})
are now $L N$- component vectors; they are conveniently
decomposed into
$N$-component vectors,
$a_i,\alpha_i$ ($i = 1. \cdots, L$) where the  $\alpha$'s are
Grassmannian
variables:
\begin{eqnarray}
     u^* &=& ( a^*_1,\cdots, a^*_L)\nonumber\\
     v^* &=& ( \alpha^*_1,\cdots,\alpha^*_L)
\end{eqnarray}
Since the matrices $C_i$($i = 1, \cdots, L$) are associated
with the
hopping
between the sites $i$  and $i+1$ , we have
\begin{eqnarray}\label{Trr}
 &&{1\over{LN}}{\rm Tr} {1\over{z - M}}=
-{i\over{L}}\int \prod_{i=1}^N da_{1i}\cdots
da^*_{Li}d\alpha_{1i}\cdots \alpha^*_{Li}\nonumber\\
&&[(a_1^*\cdot a_1) + \cdots + (a_L^*\cdot
a_L)]{\exp}(iNz[(a_1^*\cdot a_1) +
\cdots + (a^*_L\cdot a_L)\nonumber\\
&& + (\alpha_1^*\cdot \alpha_1) + \cdots + (\alpha_L^*\cdot
\alpha_L)]
 -iN[a^*_{1c}(C^{\dag}_{1})_{cd}a_{2d} + \cdots +
\alpha^*_{Lc}(C_L)_{cd}
\alpha_{1d}])\nonumber\\
\end{eqnarray}
The average over the Gaussian distribution for the matrices
$C_1,\cdots, C_L$;
\begin{equation}
   P(C) = {1\over{Z}}{\exp}(-N{\rm Tr}[C^{\dag}_1C_1 + \cdots
+ C^{\dag}_LC_L])
\end{equation}
leads to the average Green function $G(z)$. As in
(\ref{A4}), we replace the four Fermi term by an
integration over
additional variables  $\sigma_i$'s, and, as in
(\ref{sigma}),
we integrate over the anti-commuting variables:
\begin{eqnarray}\label{devt}
&&T = \int d\alpha_1 d\alpha^*_1\cdots d\alpha_L
d\alpha_L^*
{\exp}(-N[(\alpha^*_1\cdot \alpha_1)(\sigma^*_1 + \sigma_L
- iz) +
\cdots\nonumber\\
&&+ (\alpha^*_L\cdot \alpha_L)(\sigma^*_L+\sigma_{L-1}-
iz)]\nonumber\\
&& - N [ (\alpha^*_1\cdot a_1)(a^*_2\cdot \alpha_2) + \cdots
+
(\alpha^*_L\cdot a_L)(a^*_1\cdot \alpha_1) + C.C.
])\nonumber\\
&& = {\det }
\left({\matrix{S_1& |a_1><a_2|& 0&\cdots&0&|a_1><a_L|\cr
|a_2><a_1|&S_2&|a_2><a_3|&\cdots&0&0\cr
\cdots&\cdots&\cdots &\cdots &\cdots &\cdots\cr
|a_L><a_1| & 0 & 0& \cdots& |a_L><a_{L-1}|&
S_L}}\right)\nonumber\\
&&\times \left( {N\over{\pi}}\right )^{LN}
\end{eqnarray}
where
\begin{eqnarray}
     S_1 &=& \sigma^*_1 + \sigma_L - iz\nonumber\\
     S_2 &=& \sigma^*_2 + \sigma_1 - iz\nonumber\\
       &&\cdots\nonumber\\
     S_L &=& \sigma^*_L + \sigma_{L-1} - iz
\end{eqnarray}
In the calculation of the determinant, the vector $a_i$
appears
only through its squared norm denoted as $x_i = |a_i|^2$.
Then the Green function $G(z)$ becomes
\begin{eqnarray}
   && G(z) = -{i\over{L}}({N\over{\pi}})^{L}\int \prod
d^2\sigma da_ida^*_i
(\sum_{i=1}^L x_i) T(\sigma,x)\nonumber\\
   &&{\exp}(iNz\sum x_i -N [ \sum \sigma^*_i\sigma_i +
x_1x_2 +
 \cdots x_Lx_1])
\end{eqnarray}
where the factor $(N/\pi)^L$ is due to the introduction of
the $\sigma_i$'s
(\ref{sigma}).
Changing variables from $a_i$ to $x_i$ ($|a_i|^2 = x_i$),
as was done in (\ref{x1}), we have
\begin{eqnarray}\label{GG}
 &&G(z) = -{i\over{L}} ({N\over{\pi}})^{L} [{\pi^N\over{(N-
1)!}}]^L \int
\prod d^2\sigma_i\int \prod dx_i\nonumber\\
&& (x_1\cdots x_L)^{N-1}(x_1 + \cdots + x_L) T(\sigma,
x){\exp}
[iNz(x_1+ \cdots + x_L)\nonumber\\
&&-N\sum\sigma^*_i\sigma_i - N ( x_1x_2 + \cdots +
x_Lx_1)]
\end{eqnarray}
In the large $N$ limit, repeating again the argument of the
previous section,
we have a saddle-point
\begin{equation}\label{largeN2}
     (x_{i})_c = (\sigma^*_i)_c = (\sigma_i)_c = {iz + \sqrt{8 -
z^2}\over{4}}
\end{equation}
and it leads to a semi-circle law for the density of state (note
that
(\ref{largeN2})
differs from (\ref{largeN1}) since we have used in this
section a slightly
different normalization of the probability distribution).

As before the determinant $T(\sigma_c,x_c)$
vanishes at $z=0$ . We then have to expand the variables of
integration
around these saddle points.
We shift the complex variables $\sigma_{i}^* - iz/2$ to
$\sigma^*_{i}$ and
$\sigma_{i} - iz/2$ to $\sigma_{i}$. This gives
\begin{equation}
     S_1 \cdots S_L = (\sigma^*_1 + \sigma_L) \cdots
(\sigma^*_L + \sigma_{L-1})
\end{equation}
It is convenient to replace the variables $x_i$'s, which are
positive by
definition, to the variables $(\lambda_1, \lambda_2, t_3, t_4,
\cdots,
t_L)$ defined by
($i = 1, \cdots, L$),
\begin{eqnarray}
     \lambda_1 &=& x_1 + x_3 + \cdots + x_{L-1}\nonumber\\
     \lambda_2 &=& x_2 + x_4 + \cdots + x_L
\end{eqnarray}
and
\begin{eqnarray}\label{xt}
      &&x_1 = \lambda_1 ( 1 - t_3 - \cdots - t_{L-1}
)\nonumber\\
      &&x_2 = \lambda_2 ( 1 - t_4 - \cdots - t_L )\nonumber\\
      &&x_{2n-1} = \lambda_1 t_{2n-1} \hskip 10mm (n\neq
1)\nonumber\\
      &&x_{2n} = \lambda_2 t_{2n} \hskip 10mm (n\neq 1)
\end{eqnarray}
The Jacobian $J$ for this
transformation is $J= (\lambda_1 \lambda_2)^{L/2-1}$.
The Green function is then written as
\begin{eqnarray}\label{DD1}
  &&G(z) =  - {i\over{L}} {N^{2L}\pi^{L(N-1)}\over{(N!)^L}}
\int \prod d^2 \sigma_i \int d\lambda_1 d\lambda_2
dt_3\cdots dt_6
(\lambda_1\lambda_2)^{{L\over{2}}-1}\nonumber\\
&&[\lambda_1\lambda_2]^{L(N-1)\over{2}}[(1-t_3-\cdots -
t_{L-1})(1-t_4-
\cdots - t_L)t_3t_4
\cdots t_L]^{N-1}(\lambda_1 + \lambda_2) T\nonumber\\
&& {\exp}(iNz(\lambda_1 + \lambda_2 ) - N (\lambda_1
\lambda_2) h(t))\nonumber\\
&& {\exp} (-N\sum_{i=1}^L \sigma^*_i\sigma_i - i{zN\over{2}}
\sum_{i=1}^L (\sigma^*_{i}
+ \sigma_{i}))
\end{eqnarray}
in which
\begin{equation}\label{DD2}
  h(t) = t_1t_L +t_2 t_1 +
t_3 t_2 + \cdots + t_L t_{L-1}
\end{equation}
\begin{eqnarray}
     && t_1 = 1 - t_3 - t_5 - \cdots - t_{L-1}\nonumber\\
     && t_2 = 1 - t_4 - t_6 - \cdots - t_L
\end{eqnarray}
We return now to the variables $p$ and $q$ as before,
\begin{eqnarray}
      \lambda_1\lambda_2 &=& p\nonumber\\
      \lambda_1 + \lambda_2 &=& 2\sqrt{p}q,
\end{eqnarray}
with Jacobian $J= 1/\sqrt{q^2-1}$.
In the large $N$ limit, the saddle point  for the $t_i$'s  is
\begin{equation}
     t_{2n-1} = t_{2n} = {2\over{L}}
\end{equation}
and $h(t_{c}) = 4/L$.
Since $T(\sigma_c,x_c)$ is vanishing, we have to expand
$\sigma_i$ and
$p,q,t_i$ around their values at the saddle-point. First, we
integrate
over $q$ with the change of variable $ q = {\cosh} \phi$.
Then we get,
\begin{eqnarray}\label{V100}
  &&G(z) = - {2i\over{L}}{N^{2L}\pi^{L(N-1)}\over{(N!)^L}}\int
\prod d^2\sigma_i
\int dp p^{{NL\over{2}}-{1\over{2}}}\int \prod
dt_i\nonumber\\
&& [( 1 - t_3 - t_5 - \cdots - t_{L-1})(1 - t_4 - t_6 - \cdots -
t_L)t_3
t_4 t_5 \cdots t_L]^{N-1} T \nonumber\\
&&K_1(-2i\zeta \sqrt{p})\exp [ - Nph(t) - N \sum \sigma^*_i
\sigma_i
- {i\over{2}}\zeta \sum (\sigma^*_i + \sigma_i)]
\end{eqnarray}
The saddle-point is now  $t_c = {2\over{L}}$,
$p_c = {L^2\over{8}}$ and $h(t_c)= {4\over{L}}$. The
determinant $T$
vanishes at this  saddle-point. With the polar coordinate
representation for $\sigma_i$,
\begin{equation}
  \sigma_i = \sqrt{r_i}e^{i\theta_i}
\end{equation}
the saddle-point appears now  at
\begin{eqnarray}
    &&(r_i)_c = {1\over{2}}\nonumber\\
    &&(\theta_{2n-1})_c = - (\theta_{2n})_c = \theta_1
\end{eqnarray}
We denote the deviations from this saddle-point  as
$\theta'_i$,
\begin{eqnarray}
    && \theta_{2n-1} = \theta_1 + \theta'_{2n-
1}\nonumber\\
    && \theta_{2n} = - \theta_1 + \theta'_{2n}
\end{eqnarray}
with $\theta'_1 = 0$.
The determinant $T$ is independent of $\theta_1$, and thus
the integral
over $ \theta_1 $ which appears in  (\ref{V100}) is
\begin{equation}
   A = \int d\theta_1 e^{-i\zeta \sum_{i=1}^L \sqrt{r_i}
\cos (\theta_1 - (-1)^i \theta'_i)}
\end{equation}
Expanding up to second order in the $\theta'_i$'s , after
integration over
$\theta_1$, we get
\begin{equation}\label{V101}
   A \simeq 2\pi J_0(\zeta \sum_{i=1}^L \sqrt{r_i}) +
O({\theta'_i}^2)
\end{equation}
It is only the first term of (\ref{V101}) which matters, since
$T$ is
vanishing for the saddle-point,
and the fluctuations over the $\theta'_i$'s are of  higher
order.
Therefore the imaginary part of $G(z)$ is given by the
product of $J_0$ and
$J_1$ as before.
\begin{eqnarray}
 {\rm Im} G(z) &=& {2\pi^2\over{L}}{N^{2L}\pi^{L(n-1)}
\over{(N!)^L}}({1\over{2}})^L \int \prod_{i=1}^L dr_i
\prod_{j=2}^Ld\theta'_j\int dp p^{{NL-
1\over{2}}}\nonumber\\
&&\int \prod_{i=1}^L dt_i (t_1 t_2 \cdots t_L)^{N-1}\delta (
t_1 + t_3 + \cdots + t_{L-1})\nonumber\\
&&\delta(t_2 + t_4 + \cdots + t_L) T\nonumber\\
&& {\exp}(-Nph(t) - N \sum_{i=1}^L r_i) J_1(2\zeta
\sqrt{p})J_0
(\zeta\sum_{i=1}^L\sqrt{r_i})
\end{eqnarray}
The factor $({1\over{2}})^L$ comes from Jacobian of $dr$.

We still
have to integrate over $p,t_i,\theta'_i,r_i$. However, the
integration around the saddle-point for $t_i$ and $\theta'$
may be
avoided, although they could also be obtained after quite
lengthy
calculations. Indeed one may notice that if we had
calculated, instead of
the average resolvent, the average
value of 1, we would have had an  integral similar to (4.15).
If we
attempted to compute one through a  large $N$-limit
analysis,  we would
obtain again
a product  of two Bessel functions $J_0(N \mu\sum
\sqrt{r_i})$
$J_1(2 N \mu\sqrt{p})$, in which $N \mu$ is the real part of
$\zeta$.
The coefficient of this term should thus vanish.
Therefore if we replace $p$ and $r_i$ by   their  saddle-point
values, then
the contributions of the fluctuations of $\theta'_i$ and $t_i$,
which  also have a factor $J_0(L N \mu/\sqrt{2})J_1(L N
\mu/\sqrt{2})$,
should cancel each other.
 The only difference between the calculation of the average
resolvent and
of the identity,   is  a relative factor
$\sqrt{p}$.
This contribution is also exactly cancelled by those coming
from the
expansion of the Bessel function $J_1(2 N \mu\sqrt{p})$:
\begin{equation}\label{525}
J_1(2 N \mu \sqrt{p}) \simeq J_1({L N \sqrt{2}\over{2}}\mu)
+
p' {2\sqrt{2} N \mu
\over{L}}J_0({L N \mu\over{\sqrt{2}}}) - p'
{4\over{L^2}}J_1({L N \mu
\over{\sqrt{2}}})
\end{equation}
It is thus possible to replace  $t_i$ and $\theta_i$ by their
values at the
saddle-point.
The Bessel function $J_0(\zeta \sum \sqrt{r_i})$ is expanded
as
\begin{equation}
   J_0( N \mu \sum \sqrt{r_i}) \simeq J_0({L N
\mu\over{\sqrt{2}}})
- {N \mu\over{\sqrt{2}}}\sum_{i=1}^L r'_i J_1({L N
\mu\over{\sqrt{2}}})
\end{equation}
We are thus left with two terms, which are proportional to
$[J_0(L N \mu/\sqrt{2})]^2$  and $[J_1(L N \mu/\sqrt{2})]^2$.
The coefficients of these therms are evaluated by
integrating over $p'$
and $r'$ .
The determinant $T$  factorizes as
\begin{equation}\label{facL}
    2^{L\over{2}}\prod_{k=1}^{L\over{2}} ( 1 -
{8\over{L^2}}p \cos^2({2\pi\over{L}}k))
\end{equation}
It is simply the product of the eigenvalues for a periodic
chain
with  nearest neighbor interactions.
Setting $p = {L^2\over{8}} + p'$,
and expanding this product up to order $p'$, we find
\begin{eqnarray}\label{528}
   && - p' ({8\over{L^2}})
2^{{L\over{2}}}\prod_{k=1}^{{L\over{2}}-1}
( \sin^2({2\pi\over{L}}k) -
{8\over{L^2}}p'\cos^2({2\pi\over{L}}k))\nonumber\\
&&\simeq - p'
({8\over{L^2}})2^{{L\over{2}}}\prod_{k=1}^{{L\over{2}}-1}
\sin^2({2\pi\over{L}}k)\nonumber\\
&& = - p' 2^{3 - {L\over{2}}}
\end{eqnarray}
where we have used the identity:
\begin{equation}\label{ident}
\prod_{k=1}^{n-1} \sin ({\pi\over{n}}k) =
{n\over{2^{n-1}}}
\end{equation}
The $\sigma$ integral, which appears in this calculation,
may be done exactly.
After integration over the $\sigma_{i}$'s, we have
\begin{eqnarray}\label{INN}
 &&I_{NL} = \int \prod_{i=1}^{L} d^2 \sigma_i [( \sigma^*_1 +
\sigma_L) (
\sigma^*_2 + \sigma_1) \cdots ( \sigma^*_L + \sigma_{L-1}
)]^N
\exp (- N \sum \sigma^*_i \sigma_i )\nonumber\\
&& = ({\pi\over{N^{{N\over{2}}+1}}})^L C(N,L)(
N!)^{L\over{2}}
\end{eqnarray}
The quantity $C(N,2k)$ is expressible as an integral over
angles; for instance  in the $k=3$ case,
 it reads
\begin{equation}\label{angularint}
    C(N,6) = {1\over{(2\pi)^2}}\int_{0}^{2\pi}
\int_{0}^{2\pi} d\theta_1 d\theta_2[ 1 + e^{i\theta_1}]^N [ 1 +
e^{-i\theta_1
-i\theta_2}]^N[1 + e^{i\theta_2}]^N
\end{equation}
In the large $N$ limit, we exponentiate the integrand of
(\ref{angularint})
and expand the $\theta$'s  around $\theta = 0$.
Then we have  $ 2 (8^N)/(\pi\sqrt{3}N)$.
For  general $k$,
 we get
\begin{equation}
  C(N,2k) \simeq {2^{kN}\over{(2\pi)^{k-1}}}({\pi\over{N}})^
{{k-1}\over{2}}{2^{k-1}\over{\sqrt{\prod_{n=1}^{k-1}(1 +
\cos(
{\pi n\over{k}} ) )}}}
\end{equation}
 From the identity,
\begin{equation}
    \prod_{r=1}^{l-1} \cos ({\pi r\over{2l}}) =
{\sqrt{l}\over{2^{l-1}}}
\end{equation}
and setting $k = {L\over{2}}$, we have
\begin{eqnarray}
   &&\prod_{n=1}^{{L\over{2}}-1} ( 1 + \cos ({2\pi
n\over{L}}))
   = 2^{{L\over{2}}-1} \prod_{n=1}^{{L\over{2}}-1}
 \cos^2({\pi n\over{L}})\nonumber\\
   && = 2^{-{L\over{2}}}L
\end{eqnarray}
Thus we get
\begin{equation}
   C(N,L) \simeq {2^{LN\over{2}}
\over{\pi^
{{L\over{4}}-{1\over{2}}}}}
{2^{L\over{4}}\over{\sqrt{L}}}{1\over{N^{{L\over{4}}-
{1\over{2}}}}}
\end{equation}
and
$I_{NL}$ in ({\ref{INN}}) reads, for the case of an arbitrary
$L$ ,
\begin{equation}\label{536}
I_{NL} \simeq {\pi^{L+{1\over{2}}}\over{N^{L-{1\over{2}}}}}
{2^{{L\over{2}}(N+1)}\over{\sqrt{L}}}e^{-{L\over{2}}N}
\end{equation}

The remaining  integration over $p$ may again be done by
the
saddle point
method;
setting $p = {L^2\over{8}}
+ p'$, we have
\begin{eqnarray}\label{537}
&&I_p=\int dp p^{{NL\over{2}}-{1\over{2}}} p'^2 e^{-
{4\over{L}}Np}
\simeq ({L^2\over{8}})^{{NL\over{2}}-{1\over{2}}}e^{-
{LN\over{2}}}
\int dp' p'^2 e^{- {16\over{L^3}}Np'^2}\nonumber\\
&&\simeq {\sqrt{2\pi}\over{64N^{3\over{2}}
}}({L^2\over{8}})^{NL\over{2}}L^{7\over{2}}
e^{-{LN\over{2}}}
\end{eqnarray}
Thus the piece of the imaginary part of $G$, which is
proportional to
$J^2_0({L\sqrt{2}
\over{2}}\zeta)$,  is obtained by mutiplying together the
coefficients of
(\ref{525}),(\ref{528}),(\ref{536}),(\ref{537}) and the
contribution coming from the integrals over the $t_i$'s ,
which is given in
an appendix:
\begin{eqnarray}
&&({\rm Im} G)_{2a} \simeq
 - {L\over{2}}({L^L\over{4^L}}) \tilde D \pi N \mu J^2_0
({L\sqrt{2}\over{2}} N \mu)\nonumber\\
&& = - {L\over{8}}\pi N \mu J^2_0({L\sqrt{2}\over{2}}N
\mu)
\end{eqnarray}
where $N \mu$ is a real part of $\zeta$; the calculation of
$\tilde D$ is given in Appendix A, in which it is shown that
$\tilde D =
4^{L-1}/L^L$.
We have also the same coefficient for the other piece of the
imaginary part
which is proportional to $J_1^2(LN\mu \sqrt{2}/2)$.
Thus we obtain for this one-dimensional ring,
\begin{equation}
   \rho(\mu) = {L N \mu\over{8}} [ J^2_0({L N
\sqrt{2}\over{2}}\mu) +
J^2_1({L N \sqrt{2}\over{2}}\mu) ]
\end{equation}
If  we replace $LN\mu/\sqrt{2}$ by $2 x$, then this equation
becomes
$\rho (\mu)= {x\over{2\sqrt{2}}}[J_0^2(2x) + J_1^2(2x)]$; it
is identical
to (\ref{crossover}) (except  for a  trivial factor $1/\sqrt{2}$
which
comes from a different normalization in the probability
distribution in
this section; for that reason the edge  of the density of state
is  now at
$z_c = 2
\sqrt{2}$ instead of $2$).
Thus we have obtained a  behavior at the origin for a chain
of $L$ matrices
which, up to a normalization, is identical to the previous
simple case.

\sect{Lattice of matrices}

  If we now consider a higher dimensional  lattice, with a
total number of
lattice points equal to  $L$, and
periodic boundary conditions, one may ask again the same
question.
As before, we take an $LN\times LN$ random matrix , with
$N\times N$ block
elements, corresponding to the hopping between  nearest
neighbours on a
lattice. Non-neighbouring sites are not coupled and are
represented in the
total matrix by block matrices of zeroes.
Generalizing the expression of ({\ref{Trr}}), we have
\begin{eqnarray}\label{Trr2}
    &&{1\over{LN}}{\rm Tr}{1\over{z - M}}= - {i\over{L}}
\int \prod_{i=1}^{N}da_{1i}\cdots da^*_{Li}d\alpha_{1i}\cdots
d\alpha^*_{Li}\nonumber\\
&&[(a^*_1\cdot a_1)+ \cdots + (a^*_L\cdot a_L)]{\exp}
(iNz[(a^*_1\cdot a_1)
+ \cdots + (a^*_L\cdot a_L)\nonumber\\
&&+ (\alpha^*_1\cdot \alpha_1) + \cdots + (\alpha^*_L\cdot
\alpha_L)] - iN[
a^*_1 (C^{\dag}_{1,2})a_2 + \cdots ])
\end{eqnarray}
The last term of ({\ref{Trr2}}) reproduces the connectivity
of the lattice.

Integrating over the random complex matrix $C_{i,j}$, and
using the
$\sigma_{i,j}$ variables
of ({\ref{sigma}}), we have, as in  ({\ref{GG}}),
\begin{eqnarray}
   &&G(z) = - {i\over{L}}({N\over{\pi}})^L[{\pi^N\over{(N-
1)!}}]^L \int
     \prod_{<i,j>} d^2 \sigma_{i,j} \int \prod dx_i\nonumber\\
   &&(x_1 \cdots x_L)^{N-1}( x_1 + \cdots + x_L ) T(\sigma, x)
{\exp} [iNz(
x_1 + \cdots + x_L )\nonumber\\
&& -N\sum_{<i,j>} \sigma^*_{i,j} \sigma_{i,j} - N ( x_i M_{ij} x_j
)]
\end{eqnarray}
where we used the notation $|a_i|^2 = x_i$ and $<i,j>$ are a
pair of
nearest neighbours; $M_{ij}$ is the
adjacency matrix of the lattice.
One can use a method similar to that of the one dimensional
chain; we divide
the lattice points into two groups, with odd and even indices.
This change of variable leads to an expression similar
to ({\ref{D1}})
\begin{eqnarray}\label{GNN}
   &&G(z) = - {i\over{L}}{N^{2L}\pi^{L(N-1)}\over{(N!)^L}} \int
d^2
\sigma_i \int d\lambda_1 d\lambda_2 dt_1 \cdots dt_L
(\lambda_1
\lambda_2)^{{L\over{2}}-1}\nonumber\\
&&(\lambda_1 \lambda_2)^{{L(N-1)\over{2}}}[t_1 t_2 t_3
\cdots t_L]^{N-1}
(\lambda_1 + \lambda_2) T\nonumber\\
&&\delta(t_1 + t_3 + \cdots + t_{L-1} - 1) \delta(t_2 + t_4 +
\cdots
+ t_L - 1 )\nonumber\\
&& {\exp}( iNz (\lambda_1 + \lambda_2) - N \lambda_1
\lambda_2 h(t)
 - N \sum_{<i,j>} \sigma^*_{i,j} \sigma_{i,j})
\end{eqnarray}
where $h(t) = t_i M_{ij} t_j$. By the change of variables,
$\lambda_1
\lambda_2 = p$ and $\lambda_1 + \lambda_2 = 2 \sqrt{p} q$,
we integrate
over q, and obtain immediately
$K_1(-2i\zeta\sqrt{p})$ as ({\ref{K1}}). We replace
$Nz$ by $\zeta$. The saddle-point is $(t_i)_c = {2\over{L}}$,
$ \lambda_c = {L\over{4}}$, $h(t_c) = {8\over{L}}$ and $p_c =
{L^2\over{16}}$.
(Note a factor of two in the normalization compared to the
one-dimensional
problem).
We shift the complex variable $\sigma^*_{i,j} - iz/4$ to
$\sigma^*_{i,j}$
and $\sigma_{i,j} - iz/4$ to $\sigma_{i,j}$.
This shift gives an extra term to (\ref{GNN}) of $\exp ( - izN
\sum (\sigma^*_{
i,j} + \sigma_{i,j} )/4 )$.
 We write $\sigma_{i,j}$ as
\begin{equation}
     \sigma_{i,j} = \sqrt{r_{i,j}} e^{i\theta_{i,j}}
\end{equation}
We expand the variable $\theta$ around the saddle point
$\theta_{i,j}$,
\begin{eqnarray}
       &&\theta_{2i-1,j} = \theta_{1,2} + \theta'_{2i-1,j}
\nonumber\\
       &&\theta_{2i,j} = - \theta_{1,2} + \theta'_{2i,j}
\end{eqnarray}
As in the previous cases, we may simplify
\begin{equation}
     {\exp} (-{i\over{4}}\zeta \sum (\sigma^*_{i,j} +
\sigma_{i,j}))
\simeq {\exp} ( - {i\over{2}} \zeta \sum \sqrt{r_{i,j}}\cos
\theta_{1,2})
\end{equation}
The deviations from the saddle-point  $\pm \theta_{1,2}$
may be
neglected since $T$ is vanishing at the saddle-point.
Therefore, by  integration over $\theta_{1,2}$, we  obtain
again
$J_0({N \mu\over{2}}\sum \sqrt{r_{i,j}})$, where $N \mu$ is
the real part of
$\zeta$.
We note the values at the saddle-point are $(r_{i,j})_c =
{1\over{4}}$ and
$p_c = {L^2\over{16}}$.
Thus we have the following expression:
\begin{eqnarray}\label{twoIm}
    &&{\rm Im} G = {2\pi^2\over{L}}{N^{2L}\pi^{L(N-
1)}\over{(N!)^L}}
({1\over{2}})^L \int \prod dr_{i,j} \prod d\theta'_{i,j}
  \int dp p^{NL-1\over{2}} \int \prod dt_i\nonumber\\
&& [ t_1 \cdots t_L]^{N-1}
T \delta(t_1 + t_3 + \cdots + t_{L-1} - 1)\delta ( t_2 + t_4 +
\cdots
+ t_L - 1)\nonumber\\
&& J_1(2 N \mu\sqrt{p}) J_0 ({N \mu\over{2}}\sum_{<i,j>}
\sqrt{r_{i,j}})
 e^{-Nph(t) - N \sum r_{i,j}}
\end{eqnarray}

The determinant $T$ is expanded around the saddle point. If,
for definiteness
we specialize to a two-dimensional square lattice, $T$
contains the factor:
\begin{eqnarray}\label{ffac}
   &&\prod_{k_1,k_2=1}^{\sqrt{L}}[ 1 - {2\over{L}}\sqrt{p}
{\cos} {2\pi\over{\sqrt{L}}} k_1 - {2\over{L}} \sqrt{p}
{\cos } {2\pi \over{\sqrt{L}}} k_2]\nonumber\\
&& = ( 1 - {4\over{L}}\sqrt{p})\prod_{k_1,k_2=1}^{'}[1 -
{2\over{L}}
\sqrt{p} {\cos}{2\pi\over{\sqrt{L}}} k_1 - {2\over{L}}\sqrt{p}
{\cos} {2\pi\over{\sqrt{L}}}k_2]
\end{eqnarray}

Expanding $p$ around the saddle point, $p = {L^2\over{16}}+
p'$, we find
that the
first factor of (\ref{ffac}) becomes
\begin{equation}
    1 - {4\over{L}}\sqrt{p} = - {8\over{L^2}}p' + O(p'^2)
\end{equation}

We also have to expand the Bessel functions and we obtain
\begin{equation}
   J_1(2 N \mu \sqrt{p}) = J_1({L N \mu\over{2}}) +
 {4N \mu\over{L}}p' J_0({L N \mu\over{2}}) -
{8\over{L^2}}p'J_1({L N \mu\over{2}})
\end{equation}
The last term of order $p'$ cancels as before with the term
of
$1/\sqrt{p}$ in (\ref{twoIm}), as seen from the expansion
\begin{equation}
\sqrt{p} \simeq {L\over{4}}( 1 + {8\over{L^2}}p')
\end{equation}
Repeating the procedure used for the one-dimensional
chain,
we obtain an identical form for the density of state near the
origin,
\begin{equation}
    \rho(\mu)  = C\mu ( J_0^2({L N \over{2}}\mu) + J_1^2({L
N \over{2}}\mu) )
\end{equation}
where the coefficient $C$ is ${L N \over{16}}$. For this
two-dimensional lattice,
the end point of the semi-circle density of state is, with our
normalizations,  $\mu = 4$; this is why we have an extra
factor of one half
in our scaling variable compared to  (\ref{crossover}).

\sect{ Non-Gaussian probability distribution}
\vskip 5mm
  Up to now, we have considered a Gaussian distribution for
the random
matrix $C$ (\ref{Gaussian}). If one modifies this distribution,
for instance
by adding quartic terms in the exponential, the average
density of
eigenvalues will no longer obey a semi-circle law. However
earlier studies
on the usual unitary ensemble, revealed that the  behavior
near the edge of
the semi-circle was not affected by non-Gaussian terms.
The cross-over
there is always given in terms of Airy functions
(\ref{edge Airy}). We are
thus led to investigate whether this universality is also
valid for the
oscillation of the density of state near the origin that we
have found for
the block-matrices that we are considering in this article.

Let us consider a non-Gaussian distribution, for the simple
one-matrix
(i.e. $L=2$-model) of section 3, but it will be clear that the
proof of
universality will apply to any lattice of matrices. Consider
for instance
the following $P(C)$:
\begin{equation}\label{NG100}
  P(C) = {1\over{Z}} \exp ( - N {\rm Tr} C^{\dag} C - Ng {\rm
Tr}( C^{\dag}
C C^{\dag}
C )).
\end{equation}
Note that the factors of $N$ in the exponential are such that
the average
correlation functions of the eigenvalues, and in particular
the density,
have a finite limit when $N$ goes to infinity. Had we put a
higher power of
$N$ in front of the quartic term, then it would not be the
case; with a
lower power of $N$, it would not contribute at all.
Then in the cross-over region of size $1\over{N}$ near the
origin, one
could argue at first sight that the quartic terms modify
simply the overall
scale,
but not the oscillatory behavior. Indeed let us write the
average density
\begin{equation}
    \rho_0(r) = <{1\over{N}} {\rm Tr}  \delta(r- C^{\dag}C)>
\end{equation}
We then integrate over the unitary group and scale $r$, as
well as the
eigenvalues of $C^{\dag}C$ by $1\over{N^2}$ . Then one sees
immediately
that the quartic terms of (\ref{NG100}) give correction of
order
$1\over{N^2}$ . However the overall normalisation remains
$g$-dependent and
therefore it modifies the scale of the cross-over function by
a factor
which is the ratio of the partition functions for the $g\neq0$
problem ,
and the Gaussian one.

But this simple-minded analysis is based on letting $N$ go to
infinity first,
and in fact it is slightly misleading. As will be seen now, the
question is
more subtle, and the non-Gaussian terms affect more than
simply the overall
normalization of the density, as we pretended in the
previous argument. It
has already been found in similar problems
 \cite{BZ1},  that by letting $N$ go to infinity  first, one
computes
only a smoothed average of the correlation function and
this is not what we
are considering. Indeed if we smoothed out the oscillatory
part of the
density near the origin, the simple universality claimed
above would be
true : the non-Gaussian part would change only the
normalization.
However, since we are interested in those oscillations, the
previous
argument is not sufficient, and we shall argue now that the
non-Gaussian
terms do modify the period of these oscillations. This change
of the
approximate period of oscillations, is in fact expected; indeed
there are
$N$ eigenvalues  distributed between
zero and the endpoint. There are thus $N$ oscillations in the
density.
If the normalization is changed,
and  say the value of $\rho(r)$ at $r=0$  multiplied by a
factor $c$, then the
approximate period of
oscillations  has to be divided by $c$.

This may be checked explicitly by returning again to the
formulation in
terms of contour integrals,  which we have developed
in the section 3; we shall apply it now perturbatively for
this
non-Gaussian distribution.

For the non-Gaussian distribution, $U_A(t)$ in (3.2) is given
by
\begin{equation}
  U_A(t,g) = {\int e^{-N\sum (r_i + r_i a_i + g r_i^2)}
\prod_{i<j}
(r_i - r_j) ({1\over{N}}\sum_\alpha^N e^{itr_\alpha})\prod
dr_i
\over{\int e^{-N\sum(r_i + r_i a_i + g r_i^2)} \prod_{i<j}(r_i -
r_j)
\prod dr_i}}
\end{equation}
This is expressible as
\begin{equation}
      U_A(t,g) = { e^{-{1\over{N}} g \sum
{\partial^2\over{\partial {a_i}^2}}}
      F\over{e^{-{1\over{N}}g \sum {\partial^2 \over{\partial
{a_i}^2}}}
         D }}
\end{equation}
where
\begin{eqnarray}
     && F = \int e^{-N\sum ( r_i + r_i a_i)}\prod_{i<j}(r_i - r_j)
({1\over{N}}
\sum_{\alpha = 1}^N e^{itr_\alpha})\prod dr_i\nonumber\\
&& D = F( t = 0 )
\end{eqnarray}
Expanding this $U_A(t,g)$ up to order $g$, we get
\begin{equation}
U_A(t,g) \simeq U_A(t,g=0) - {g\over{N}}\sum_{i=1}^N
\left [ ({\partial^2 F\over{\partial {a_i}^2}}){1\over{D}}
- ({\partial^2 D\over{\partial {a_i}^2}}) {F\over{D^2}}\right ]
+ O(g^2)
\end{equation}
Noting that
\begin{eqnarray}
{\partial U_A\over{\partial a_i}}&=&({\partial
F\over{\partial a_i}})
{1\over{D}} - {F\over{D^2}}({\partial D\over{\partial a_i}})
\nonumber\\
{\partial^2 \over{\partial {a_i}^2}} U_A(t,0)&=&
({\partial^2 F\over{\partial {a_i}^2}}){1\over{D}} -
{F\over{D^2}}
({\partial^2 D
\over{\partial {a_i}^2}}) + 2 {F\over{D^3}}({\partial
D\over{\partial a_i}})^2
- 2 ({\partial F\over{\partial a_i}})({\partial D\over{\partial
a_i}}){1\over{D^2}}\nonumber\\
{}
\end{eqnarray}
the term of order $g$, denoted by  $\delta U_A$, becomes
\begin{equation}
\delta U_A = -  {g\over{N}}\sum_{i=1}^N {\partial^2
\over{\partial {a_i}^2}}
U_A(t,0) -  {2g\over{N}} \sum_{i=1}^N ({\partial
U_A\over{\partial a_i}})
({\partial {\ln D}\over{\partial a_i}})
\end{equation}
The expression of D is given in (3.7)
and we have
\begin{equation}
    {\partial \ln D\over{\partial a_i}} = \sum_{k\neq
i}{1\over{(a_i - a_k)}}
- {N\over{1 + a_i}}
\end{equation}
Using the contour representation in (3.10), we get
\begin{eqnarray}
{\partial U_A\over{\partial a_i}}& = - {1\over{N}}\oint
{du\over{2\pi i}}
\left ({1 + u\over{1 + u - {it\over{N}}}}\right )^N {1\over{(u
- a_i)^2}}
\prod_{\gamma\neq i}\left ({ u - a_\gamma -
{it\over{N}}\over{u - a_\gamma}}
\right)
\nonumber\\
{\partial^2 \over{\partial {a_i}^2}} U_A(t,0)& =
     -{2\over{N}} \oint {du\over{2\pi i}}
( {1+u\over{ 1 + u - {it\over{N}}}})^N {1\over{
(u - a_i)^3}}\prod_{\gamma\neq i}\left( {u - a_\gamma
-{it\over{N}}\over
{u - a_\gamma}}\right )
\end{eqnarray}
Noting that
\begin{eqnarray}
    &&{1\over{(u - a_i)^2(a_i - a_k)}}( 1 - {it\over{N(u-a_k)}})
    + {1\over{( u - a_k)^2(a_k - a_i)}}( 1 - {it\over{N(u -
a_i)}})\nonumber\\
&&= {2u - a_i - a_k - {it\over{N}}\over{(u - a_i)^2 ( u-
a_k)^2}}
\end{eqnarray}
we are able to express the contribution $\delta U_A$ in the
contour
integral.
Thus we get the contour representation in the first order of
$g$;
by letting $a_i$ goes to zero,
\begin{eqnarray}
 &&\delta U_0 =  {2g\over{N}}\oint {du\over{2\pi i}} \left( {1
+ u\over{
1 + u - {it\over{N}}}}\right )^N {1\over{u^3}}\left({u -
{it\over{N}}\over
{u}}\right )^{N-1}\nonumber\\
&&+ {2g\over{N}}\oint {du\over{2\pi i}}\left ( {1 + u\over{1
+ u - {it\over
{N}}}}\right )^N  [ ({N-1\over{2}})({2u -
{it\over{N}}\over{u^4}})({u -
{it\over{N}}
\over{u}})^{N-2} \nonumber\\
&&  - {N\over{u^2}}({u - {it\over{N}}\over{u}})^{N-1}
 ]
\end{eqnarray}
Immediate checks for $N$=1, 2 and 3, and  also the large $N$
 limit, are easy and
they agree with the result known by other methods. Thus
this expression
is exact to this order in $g$.

We now consider the cross-over region near the origin;
replacing $t$ by
$N^2 t$, and $u$ by $N u$, we have
\begin{eqnarray}
\delta U_0(t) &=& 2g
        \oint {du\over{2\pi i}}\left ( { 1 + {1\over{Nu}}\over
      {1 + {1\over{N(u - it)}}}}\right )^N  [ {1\over{N^3}}
{1\over{u^2}}({1\over{u - it}})\nonumber\\
      &&+ {N-1\over{ 2 N^3}}{1\over{u^2}}{2u - it\over{(u -
it)^2}} -
 {1\over{N}}{1\over{u ( u - it)}} ]
\end{eqnarray}
Only the last term in the bracket contributes in the large N
limit. It gives,
 by the Fourier transform in (3.3),
\begin{equation}\label{delrho}
\delta \rho (\mu) = 2 g N J_{0}^2(2x)
\end{equation}
where $x = N\mu$. The overall factor due to the change of
normalization is $ 1 + 2 g$, which agrees with the expression
for the density
of state which was found in [2]. Then, up to this order, we
may interprete
this result as reading
\begin{equation}
     \rho_{0}(\mu) = C(g) {N\mu \over{2}}[ J_{0}^2( 2 N
\mu\sqrt{C(g)})
+ J_{1}^2( 2 N \mu\sqrt{ C(g)})]
\end{equation}
where $C(g)= 1 +2g +O(g^2)$. This expression makes it clear
that the
integrated density of state remains properly normalized to
one.
Indeed, expanding this expression for $g$ small, we find the
term
$g J_{1}^2( 2 x)$ cancel, and obtain (\ref{delrho}).
This result is expected; it is exactly the universality which
was claimed earlier up to this order in $g$.

We are thus tempted to conjecture that these oscillations
are indeed universal, namely that for an arbitrary
probability distribution of the form
\begin{equation}
P(C)={1\over Z}e^{-N Tr V(C^{\dagger}C)}
\label{eq:anyv}
\end{equation}
with $C$ defined over a lattice
the density of states near the origin is given by
\begin{equation}
\rho(\mu)={NLF^2\mu\over
2}[J_0^2(2NLF\mu)+J_1^2(2NLF\mu)]
\label{eq:univ}
\end{equation}
with $F(V)$ some functional of $V$.

\sect{\bf Oscillation and cross-over behavior for real
matrices}
\vskip 5mm

It has long been known that for real symmetric matrices
the relevant Jacobian involves the absolute value of the van
der Monde determinant and thus the corresponding
orthogonal polynomial analysis becomes quite complicated.
However, by using some remarkable identities, we can
actually
treat the cross-over behavior near the origin of the density
of state of a $(2 N + 1)\times (2 N + 1)$ real matrix
$M$, which is made of a rectangular $(N + 1)\times N$
matrix $C$:
\begin{equation}
   M  =\left(\matrix{0 & C^{T}\cr
                 C& 0 \cr}\right)
\end{equation}
where $C^{T}$ is the transpose of the matrix $C$. As we will
see, the
cross-over
near the origin shows a different universal behavior from
what we have studied earlier.

To understand why we treat the case of $C$ being $(N +
1)\times N$ and to get oriented, let us do a simple exercise
in power counting. Consider an $M\times N$ real matrix $C$
with its $MN$ real variables. Denote the eigenvalues
of the $N \times N$ real symmetric matrix $C^T C$ by $r_i$.
We would like to have
\begin{equation}
    dC \simeq \prod_{i<j}^N | r_i - r_j | dr_1 \cdots
dr_N
\end{equation}
To see if this is possible, let us do dimensional analysis and
count powers. The left hand side has the dimension of $r^{
{N(N-1)\over2} + N} \sim C^{N^2+N}$; on the other hand, the
right hand side has the dimension of $C^{MN}$. Equating
$MN=N^2+N$ we find $M=N+1$. This explains why we chose
$C$ to be $(N + 1)\times N$. We can of course also check this
Jacobian by direct computation, using the Fadeev-Popov
method for example.

Note that if we had chosen $C$ to be $N\times N$, we would
have
\begin{equation}
    dC \simeq \prod_{i}^{N}{1\over{\sqrt{r_i}}}\prod_{i<j}^N |
r_i - r_j
| dr_1 \cdots
dr_N
\label{sqrt}
\end{equation}
The presence of the square root factors can be deduced by
dimensional arguments or determined by a direct
computation of the Jacobian. These square roots make the
calculation much
more complicated. See below.

We have as usual the density of state
\begin{equation}
     \rho(\mu) = < {1\over{2 N + 1}}{\rm Tr} \delta (\mu - M)
>
\end{equation}
and from the block structure of M,
\begin{equation}
\rho(\mu) = < {\mu\over{N}} {\rm Tr} \delta ( \mu^2 - C^T C
) >
\end{equation}
(Note that the $(2 N + 1)\times (2 N + 1)$ real matrix
$M$ we started out with has an
eigenvalue at zero; obviously, the corresponding
eigenvector is the vector orthogonal to the $N$ columns in
the $N \times N+1$ matrix $C^T$. This eigenvalue leads to an
additonal delta function at the center of the spectrum in the
density of states.)
As in the complex matrix case, we define $\tilde \rho
(\lambda ) = <{1\over{N}}{\rm Tr}
\delta ( \lambda - C^T C )>$, with
$\rho(\mu) = \mu \tilde \rho(\mu^2)$.
With the probability distribution
\begin{equation}
    P(C) = {1\over{Z}} {\exp}( - N {\rm Tr} C^T C )
\end{equation}
this model is known as the orthogonal Laguerre ensemble
\cite {NS}.

In an obvious extension of Kazakov's method we  introduce
an external source
matrix
$A$, which we take to be an $N\times N$ Hermitian matrix
diagonalized by
the unitary
matrix $U$. The probability distribution is
\begin{equation}
     P_A(C) = {1\over{Z_A}}{\exp} ( - N {\rm Tr} C^T C - N {\rm
Tr} A C^T C )
\end{equation}
Since $C^T C$ is a real symmetric matrix, it is diagonalized
by an orthogonal
matrix $O$, and then
\begin{equation}
   {\rm Tr} A C^T C = {\rm Tr} U^{-1} \left (\matrix{a_1& & &
& \cr
        & \cdot& & & \cr
        & & \cdot & &\cr
        & & &\cdot&\cr
        & & & &a_N}\right) U O^T \left (\matrix{r_1& & & & \cr
                                                  &\cdot& & &\cr
                                                  & &\cdot& &\cr
                                                  & & &\cdot&\cr
                                                  & & & &r_N}\right) O
\end{equation}

Since there is no known analog of the Itzykson-Zuber
identity for integrating over orthogonal matrices, we would
have been stuck at this point. The crucial observation is to
notice that we could
integrate over the unitary matrix $V = U O^T$. Strictly
speaking,
$A$ is not an external
source since we integrate over all matrices unitarily
equivalent to $A$. However, since we set
the $a_j$'s to zero at the end, this procedure gives us the
correct value of $U_0(t)$. Thus, integrating
\begin{equation}
    U_A(t) \simeq \int dV \int dr_i \prod_{i<j}^N |r_i - r_j|
({1\over{N}}
\sum_{\alpha=1}^N e^{ir_\alpha}) e^{-N\sum r_i - N {\rm Tr}
V^{-1}A_{diag}V {C^T C}_{diag}}
\end{equation}
using the Itzykson-Zuber identity, we obtain
\begin{eqnarray}
U_A(t) &=& {1\over N} \sum_{\alpha=1}^{N}
\int dr_1\cdots dr_N e^{it r_{\alpha}}{\prod_{i<j}|r_i -
r_j|\over
{\prod_{i<j}(r_i - r_j)}}
\nonumber\\
&\times &\exp
(-N\sum_{i=1}^N r_i -
N\sum_{j = 1}^{N} r_{j} a_{j}).
\end{eqnarray}
where $U_A(t)$ is  normalized as $U_A(0) = 1$.

A remarkable identity
\begin{eqnarray}
  &&\int_0^{+\infty}\cdots \int_0^{+\infty}\prod_{i=1}^N
 dr_i \prod_{i<j}^N{sign{r_i - r_j}}e^{-\sum r_i b_i}
   \nonumber\\
 && = {1\over{\prod_{i=1}^N b_i}}\prod_{i<j}^N \left ({b_i -
b_j\over{b_i
 + b_j}}\right )
\end{eqnarray}
allows us to integrate over the $r_i$'s.
Let us sketch a proof of this identity. The integration region
in (8.11) can be
divided into $N!$ regions inside each of which the $r$'s are
ordered. Thus,
the integral above is equal to
\begin{equation}
  \int_0^{+\infty}\cdots \int_0^{+\infty}\prod_{i=1}^N
 dr_i  \theta(r_1 \leq r_2 \leq r_3 \leq ..... \leq r_N) e^{-\sum
r_i b_i}
\end{equation}
plus $N!-1$ similar integrals with the $r_i$'s permuted and
with a suitable sign given by $\prod_{i<j}^N{sign{r_i - r_j}}$.
(We will not be
concerned with the overall multiplicative factor in what
follows since it
is irrelevant to our calculation.) To get oriented, let us do
the
$N=3$ case. Change variables by $r_1=x_1, r_2=r_1+x_2,
r_3=r_2+x_3,.....$ to take care of the ordering. We can then
immediately do the integral above to obtain
\begin{equation}
1\over b_3(b_3+b_2)(b_3+b_2+b_1)
  \end{equation}
We now add the $5$ other terms and collect denominators;
the sum evidently
has the form
\begin{equation}
f(b) \over
b_1 b_2 b_3 (b_1+b_2)(b_2+b_3)(b_3+b_1)(b_1+b_2+b_3)
  \end{equation}
with some numerator $f(b)$ which we can now determine
by general arguments. Since our
integral vanishes whenever any two of the $b_i$'s are
equal, we
must have $f(b)=\Delta(b) P(b)$ where
$\Delta(b)=\prod_{i<j}^N (b_i-b_j)$ is
the usual van der Monde determinant and where the
polynomial $P(b)$ must be symmetric and by dimension
coutning must be of degree . Thus, $P(b)$ is
uniquely determined to be $b_1+b_2+b_3$. We have proved
the identity for
$N=3$. We then proceed by induction. Assume that the
identity has been proved for some $N$. Doing the integral
for $N+1$ along the
line described above we encounter after the first step
\begin{equation}
1\over b_1(b_1+b_2)(b_1+b_2+b_3)
....(b_1+b_2+...+b_N)(b_1+b_2+...+b_N+b_{N+1})
  \end{equation}
plus permutations.  We now first add the terms obtained by
permuting $b_1, b_2, ....., b_N$,  holding $b_{N+1}$ fixed. Then
by the inductive
process we obtain
\begin{equation}
 {1\over \prod_{i=1}^N b_i}{\Delta_N(b)\over \prod_{i < j \leq
N}
(b_i+b_j)}
{1\over (b_1+b_2+...+b_N+b_{N+1})}
  \end{equation}
plus terms obtained by interchanging $b_{N+1}$ with one of
the $b_i$'s. The subscript on $\Delta_N$
indicates that it is the van der Monde determinant for the
first $N$
$b_i$'s. Collecting common denominators and reasoning as
before, we find that
the sum is equal to
\begin{equation}
 {1\over \prod_{i=1}^{N+1} b_i}{\Delta_{N+1}(b)\over
\prod_{i < j \leq N+1}
(b_i+b_j)}
{P(b)\over (b_1+b_2+...+b_N+b_{N+1})}
  \end{equation}
By symmetry and by dimensional analysis the symmetric
polynomial $P$ must
be of degree 1 and hence equal to
$(b_1+b_2+...+b_N+b_{N+1})$. We have thus
proved the identity.
Note that if a square root factor were present, as in
(\ref{sqrt}), we
would not have this identity and thus would have difficulty
proceeding farther.

Since $b_i$  is given by
\begin{equation}
    b_i = N ( 1 + a_i - {i t\over{N}}\delta_{\alpha,i} )
\end{equation}
we have
\begin{equation}
   U_A(t) = {1\over{N}}\sum_{\alpha=1}^N ({1 +
a_{\alpha}\over{1 + a_{\alpha}
- {it \over{N}}}})\prod_{\gamma\neq \alpha} ({a_\alpha -
a_\gamma - {i t\over
{N}}\over{2 + a_\alpha + a_\gamma - {i t\over{N}}}})
({2 + a_\alpha + a_\gamma\over{a_\alpha - a_\gamma}})
\end{equation}
This is expressible in a contour representation:
\begin{eqnarray}
&& U_A(t) = - {1\over{i t}} \oint {du\over{2\pi i}} ({1 +
u\over{1 + u - {i t
\over{N}}}})\prod_{\gamma=1}^N ({2 + u + a_\gamma
\over{2 + u + a_\gamma
- {i t\over{N}}}})\nonumber\\
&& \times ({2 + 2 u  - {i t\over{N}}\over{2 + 2 u}})
\prod_{\gamma'=1}^N
({u - a_{\gamma'} - { i t\over{N}}\over{ u - a_{\gamma'}}})
\end{eqnarray}
Letting all $a_\gamma$'s go to zero, we have
\begin{eqnarray}
&& U_0(t) = - {1\over{i t}} \oint {du\over{2\pi i}} ({1 +
u\over{1 + u - {i t
\over{N}}}}) ({2 + u  \over{2 + u
- {i t\over{N}}}})^N\nonumber\\
&& \times ({2 + 2 u  - {i t\over{N}}\over{2 + 2 u}})
({u  - { i t\over{N}}\over{ u }})^N
\end{eqnarray}
where the contour is chosen around $u=0$.
We are able to check this formula by calculating the Fourier
transform,
\begin{equation}
  \tilde \rho(r) = \int_{-\infty}^{\infty} {dt\over{2\pi}} e^{-
itr}U_0 (t).
\end{equation} and verifying that we obtain the same result
for $N=2$ and  $3$ as we would have obtained by directly
integrating
\begin{equation}
    \tilde \rho(r_1) = \int \prod_{i<j} | r_i - r_j | {\exp}( - N
\sum r_i)
dr_2 \cdots dr_N
\end{equation}
We obtain the semi-circle law just as in the complex
matrix case.

Next, we study the cross-over behavior near the center of
the spectrum. Changing variables $u\rightarrow Nu$ and
$t\rightarrow N^2 t$ we obtain in the
large $N$ limit,
\begin{equation}
 U_0(t)\simeq - {N\over{2 i t}}\oint {du\over{2 \pi i}}
e^{{2\over{u}} - {2\over{
u - it}}} ( 1 + {u\over{u - it}} )
\end{equation}
By following the similar procedure as (\ref{process1}) and
(\ref{process2}), We find
\begin{equation}
  {d\over{d x^2}}\tilde\rho( x ) = - {1\over{x^2}}J_1(  x)^2 +
{1\over{x}} J_0(x) J_2(x)
\end{equation}
where $x = \sqrt{2}N \lambda$.
Thus we obtain
\begin{equation}
   \tilde\rho(\lambda) = J_0(\sqrt{2}N \lambda)^2 +
J_1(\sqrt{2}N \lambda)^2 -
{1\over{\sqrt{2}N
\lambda}}J_0(\sqrt{2}N\lambda)J_1(\sqrt{2}N\lambda)
\end{equation}
The density of state $\rho(\lambda)$ is obtained by
$\rho(\lambda) = \sqrt{2}
N \lambda \tilde\rho(\lambda)$.
The behavior is different from the complex case. The
oscillations here are
 milder than the complex case.
\sect{Discussion}

   We have explored the cross-over behavior near the origin
for the
density of state for hermitean and for real matrices made of
blocks. Although the result has been known
\cite{HZ,NS,Forr,AST}
for the one matrix model, our
derivation, in particular our derivation using Kazakov's
method, provides new and
simple expressions. We have also extended the discussion to
rings and lattices of matrices. We have proved the
universality of this cross-over behavior to first order in the
deviation from a  Gaussian distributions,  and conjecture
that this universality should hold in general.

We note also that in Kazakov's method we encounter
expressions which are
valid for a
non-vanishing external source matrix $A$.
Instead of letting all the eigenvalues of $A$ go to zero, we
may consider some specific choices of
 the eigenvalues of A. Let us, for instance, consider
$a_\gamma =
\cos ( 2\pi \gamma / N)$. Then  the contour integral
$U_A(t)$ in (\ref{114}) depends upon these $a_\gamma$'s.
However in the cross-over region, these $a_\gamma$'s may
be neglected with
respect to the variable  $u$, which was scaled by a factor $N$
in the
integral. Thus such non-zero $a_\gamma$'s are irrelevant,
and again we
would obtain with them the same universal form in the
large $N$ limit.
Unfortunately we have not been able to extend  Kazakov's
method  to the
case of a lattice of matrices.

For the lattice of matrices, we could have considered also  a
representation of the Hamiltonian as a
block matrix $M$ of the form given in (\ref{C1}). Indeed, for
a
bi-partite lattice with an even number $N$ of sites, we could
divide the lattice into
two sub-lattices A and B such that  nearest neighbors
belong to different
sub-lattices. The sub-lattice A contains sites labelled from 1
to
N/2 and the sub-lattice B from
N/2 + 1 to N.  With such a labelling the Hamiltonian $M$  for
a particle hopping on this lattice would
have a form like  (\ref{C1}).
However, the corresponding block matrix $C$ would be
sparse, containing many zero matrix
elements. Indeed, the matrix $C$
would be ${N\over{2}}\times
{N\over{2}}$, i.e. $C$ would contain  $N^2/4$ elements.  For a
$D$-dimensional
hypercubic lattice the number of
bonds in the lattice is $DN$, where $D$ is the spatial
dimension.  Then the ratio
of the number of bonds to the number of matrix elements of
$C$ is given by  $4D/N$.
When $D = N/4$, the model reduces to the one matrix model.
The example $N=4$
in the one-dimensional ring has been mentioned before.
 From this
consideration, in the large $D$ limit, if we keep $D = N/4$, we
have
a simple one matrix model of the form in (\ref{C1}) and the
universality of the cross-over behavior near the center of
the spectrum should hold.

As already remarked in \cite{HZ}, one particular simple
way to test for
the universal oscillation studied here is to compare the
height of the
first peak in the density of states to the height of the second
peak.
According (\ref{eq:univ}), the density of states is
proportional to the
universal function
\begin{equation}
r(y)=y(J_0^2(y)+J_1^2(y))
\label{eq:universal}
\end{equation}
where $y$ is the energy multiplied by some suitable
constant. The positions
of the peaks (and valleys) are determined by
\begin{equation}
{dr(y)\over dy}=J_0^2(y)-J_1^2(y)=0
\label{eq:peaks}
\end{equation}
We find for example that the ratio of the height of the first
peak to the
height of the second peak is given by 1.218. (The ratio of the
height of
the first peak to the height of the first valley is 1.58.)

A possible application is to the problem of a single particle
propagating
on a square lattice penetrated by random magnetic flux
\cite{pryor}, a
problem that has recently attracted considerable attention
\cite{others}.
Already the authors of \cite{pryor} (see figure 1 in this
reference)
noted that the density of states exhibits oscillations for finite
$N$.  It
is far from clear that our present work can be applied to this
problem
since, as we have just explained, the relevant matrix $C$ in
this random
flux problem is sparse, with the ratio of non-zero matrix
elements to the
total number matrix elements given by $16/N$.
Nevertheless, we note that a
recent numerical study of the random flux problem by
Avishai and Kohmoto
\cite{avish} found that the ratio of the height of the first
peak to the
height of the second peak was given by 1.25. We do not
know whether the
difference between 1.25 and 1.218 is real or due to
numerical
uncertainties.

\vskip 10mm
\begin{center}
{\bf ACKNOWLEDGEMENTS}
\end{center}

  We are thankful for the support by the cooperative
research project between the
Centre National de la Recherche Scientifique and the Japan
Society for the
Promotion of Science. The work of A. Z. is supported in part
by the National
Science Foundation under Grant No. PHY89-04035. He is
grateful to the
\'Ecole Normale Sup\'erieure, where part of this work was
done, for its
hospitality.

\newpage
\setcounter{equation}{0}
\renewcommand{\theequation}{A.\arabic{equation}}
{\bf APPENDIX A: {INTEGRATION FOR GENERAL L }}
\vskip 5mm

 We evaluate the following integral $D_0$ for  general  L,
which appeared in (\ref{V100}),
\begin{eqnarray}\label{diag}
  D_0 &=& \int_0^{\infty} \prod_{i=1}^{L} dt_i [ t_1 t_2 \cdots
t_L ]^{N-1}
e^{-N{L^2\over 8}h(t)}\delta(t_1 + t_3 + \cdots + t_{L-1} -
1)\nonumber\\
&& \delta(t_2 + t_4 + \cdots + t_L - 1)\nonumber\\
&=& {1\over{(2\pi)^2}}\int_{-\infty}^{\infty}
dk_1 dk_2 \int\prod dt_i [t_1 t_2 \cdots t_L]^{N-1}
e^{-Nph(t)}\nonumber\\
&& e^{ik_1(t_1 + t_3 + \cdots + t_{L-1} - 1) + ik_2 (t_2 + t_4 +
\cdots
+ t_L - 1 ) }\nonumber\\
&\simeq& {1\over{(2\pi)^2}}e^{-
{LN\over{2}}}({2\over{L}})^{L(N-1)}
\int e^{-{L^2\over{8}}N [ \sum
t'^2_i + t'_1 t'_2 + \cdots + t'_L t'_1 ]}\nonumber\\
&& e^{ik_1( t'_1 + \cdots + t'_{L-1} ) + ik_2 (t'_2 + \cdots +
t'_L)}
\prod_{i=1}^{L} dt'_i dk_1 dk_2\nonumber\\
\end{eqnarray}
We can interpret the quadratic form in the square bracket
as $t'Ht$ with $H$ the quantum Hamiltonian of a particle
hopping on  a ring (with a trivial constant site energy.) We
diagonalize $H$ and obtain for its eigenvalues the Bloch
energies $cos {2\pi k \over L}$.

 For example,
in the case $L=4$, we have
\begin{eqnarray}
D_0 &=& {1\over{(2\pi)^2}}({1\over{2}})^{4(N-1)}e^{-
2N}\int e^{-2N
(x^2_2 + x^2_3
+ 2 x^2_4 )} e^{ik_1(x_1 + x_4)+ i k_2 (-x_1 +
x_4)}\nonumber\\
&& dk_1 dk_2 dx_1
dx_2 dx_3 dx_4
\nonumber\\
&=& ({1\over{2}})^{4(N-1)}e^{-2N}\int dx_2 dx_3 dx_4
\delta(2x_4) e^{-2N(x^2_2 + x^2_3 + 2x^2_4 )}
\nonumber\\
&=& {\pi\over{4N}}({1\over{2}})^{4(N-1)} e^{-2N}
\end{eqnarray}
where $x_1 = {1\over{2}}(t_1 - t_2 + t_3 - t_4)$, $x_2 =
{1\over{\sqrt{2}}}
(t_1 - t_3)$, $x_3 = {1\over{\sqrt{2}}}(t_2 - t_4)$ and $x_4=
{1\over{2}}
(t_1 + t_2 + t_3 + t_4)$. For general $L$,  we obtain
\begin{equation}
  D_0 = ({2\over{L}})^{L(N-1)} e^{-{L\over{2}}N}
({8\pi\over{L^2 N}})^{{L-2
\over{2}}}\prod_{k=1,k\neq {L\over{2}}}^{L-1}( 1 - \cos
({2\pi\over{L}}k) )^{-{
1\over{2}}} {2\over{L}}
\end{equation}
Using the identity of ({\ref{ident}}), we have
\begin{eqnarray}
    &&\prod_{k=1,k\neq {L\over{2}}}^{L-1} (1 - \cos
{2\pi\over{L}}k ) =
 {1\over{2}}\prod_{k=1}^{L-1} \sin^2 ({\pi\over {L}}k) 2^{L-
1}\nonumber\\
&& = {L^2\over{2^L}}
\end{eqnarray}
Thus we get
\begin{equation}
   D_0 = ({2\over{L}})^{L(N-1)} e^{-
{L\over{2}}N}({\pi\over{N}})^{L-2\over{2}}
\tilde D
\end{equation}
with
\begin{equation}
\tilde D = {4^{L-1}\over{L^L}}
\end{equation}


\end{document}